\documentclass[11pt]{article}
\pdfoutput=1
\usepackage{graphicx}
\graphicspath{{./figures/}}
\usepackage{appendix}
\usepackage{latexsym,amsmath,amsfonts,amssymb,booktabs}
\usepackage[font=small]{caption}
\usepackage{slashed,upgreek,amscd,cancel,tensor,color}
\usepackage{adjustbox}
\usepackage{soul}
\usepackage[normalem]{ulem}
\usepackage[numbers,compress,square]{natbib}
\usepackage{epsfig,latexsym}
\usepackage{amsmath}
\usepackage[pdfencoding=auto]{hyperref} 
\usepackage{url}
\numberwithin{equation}{section}
\usepackage{doi}
\usepackage{subcaption}
\definecolor{MyBlue}{rgb}{0.15,0.15,0.70}
\definecolor{linkblue}{rgb}{0,0,0.8}
\definecolor{linkgreen}{rgb}{0,0.5,0}

\input epsf
\usepackage{amssymb}
\usepackage{amsmath}
\usepackage{amsfonts}
\usepackage{upgreek}
\usepackage{latexsym}
\usepackage{stfloats}
\usepackage{afterpage}

\numberwithin{equation}{section}

\newcommand{\bea}{\begin{eqnarray}}
\newcommand{\eea}{\end{eqnarray}}
\newcommand{\be}{\begin{equation}}
\newcommand{\ee}{\end{equation}}

\newcommand{\cH}{\mathcal{H}}

\newcommand{\knl}{k_{\rm NL}}
\newcommand{\kvec}{\vec{k}}
\newcommand{\qvec}{\vec{q}}
\newcommand{\uvec}{\vec{u}}
\newcommand{\vvec}{\vec{v}}
\newcommand{\xvec}{\vec{x}}
\newcommand{\nvec}{\vec{n}}
\newcommand{\yvec}{\vec{y}}

\newcommand{\eqn}[1]{Eq.~(\ref{#1})}

\newcommand{\pd}{\partial}
\newcommand{\om}{\Omega_{\rm m}}
\newcommand{\omn}{\Omega_{\rm m,0}}

\newcommand{\vx}{\vec{x}}

\newcommand{\xfl}{\xvec_{\rm fl}}

\newcommand{\tin}{t_{\rm in}}

\newcommand{\andd}{\ , \quad \text{and}  \quad}
\newcommand{\Om}{\Omega_{\rm m}}
\newcommand{\secref}[1]{Sec.~\ref{#1}}
\newcommand{\pvec}{\vec{p}}
\newcommand{\appref}[1]{App.~\ref{#1}}
\newcommand{\figref}[1]{Fig.~\ref{#1}}

\definecolor{linkcardinal}{RGB}{140, 21, 21} 
\definecolor{linkkellygreen}{RGB}{48.6, 148.6, 89.5} 
\definecolor{linktest}{RGB}{25, 145, 73} 
\definecolor{linklightblue}{RGB}{ 40, 129, 189}

\hypersetup{
colorlinks=true,
citecolor=linktest,
linkcolor=linklightblue,
urlcolor=linklightblue
}

 \definecolor{MattOrange}{rgb}{1.0,0.4,0.2}

\newcommand{\Comment}[1]{{}}

\begin{document}
\def\thefootnote{\fnsymbol{footnote}}
\setcounter{page}{1} \baselineskip=15.5pt \thispagestyle{empty}

\vspace*{-25mm}


\begin{center}

{\Large \bf Non-local-in-time structure formation and sixth-order galaxy bias}  \\[0.7cm]
{\large   Alex Edison${}^{1,2}$, Matthew Lewandowski${}^3$, Leonardo Senatore${}^4$ \\[0.7cm]}

\end{center}

\begin{center}

\vspace{.0cm}

\begin{small}
{ \textit{  $^{1}$ The Amplitudes and Insights Group, Department of Physics \& Astronomy, \\ Northwestern University, Evanston, Illinois 60208, USA}}
\vspace{.05in}

{ \textit{  $^{2}$ Center for Interdisciplinary Exploration and Research in Astrophysics (CIERA), \\ Northwestern University, 1800 Sherman Ave, Evanston, IL 60201, USA}}
\vspace{.05in}

{ \textit{  $^{3}$ Theoretical Physics Department, CERN, 1211 Geneva, Switzerland}}
\vspace{.05in}

{ \textit{  $^{4}$ Institute for Theoretical Physics, ETH Zurich,
8093 Zurich, Switzerland}}
\vspace{.05in}

\end{small}
\end{center}

\vspace{0.5cm}

\begin{abstract}
We present a systematic construction of the non-local-in-time galaxy bias expansion in the Effective Field Theory of Large-Scale Structure.  In order to fully capture time non-locality up to sixth order, we must take into account that every field can contribute non-locally from a separate time in the past.  Starting from the general non-local-in-time expression for the galaxy overdensity, we explicitly compute the complete sixth-order basis of bias operators at leading order in spatial derivatives, finding 57 independent biases, compared to 46 in the corresponding local-in-time expansion.  As previously found at fifth order, this difference implies that higher-order clustering statistics can distinguish between local- and non-local-in-time galaxy formation, and thus are sensitive, at a single redshift, to the formation time of galaxies.  Along the way, we obtain recursion relations for the perturbative kernels and show that they satisfy specific multi-leg soft limits when the sum of a subset of the external momenta goes to zero.
\end{abstract}

\tableofcontents

\vspace{.5cm}

\def\thefootnote{\arabic{footnote}}
\setcounter{footnote}{0}

%
%
%
%

\section{Introduction, notation, and conclusion}

In the weakly non-linear regime, correlations of density fluctuations observed in the Large-Scale Structure (LSS) of the universe can be predicted perturbatively within the framework of the Effective Field Theory (EFT) of Large-Scale Structure \cite{Baumann:2010tm, Carrasco:2012cv}.  In this approach, the dynamics of long-wavelength perturbations are reproduced through a systematic expansion in both the amplitude of fluctuations and their spatial derivatives relative to the non-linear scale, which is roughly $10$~Mpc today.  The effects of short-distance physics on large-scale clustering are captured by a set of EFT coefficients, which, from the long-wavelength point of view, are a priori unknown but can be determined directly by comparison with cosmological data.  Overall, this approach ensures that our theory of gravitational clustering is consistent with the basic symmetries of General Relativity and provides a controlled framework to compute large-scale observables.  

A central ingredient in modeling LSS is the description of galaxies as biased tracers of the underlying dark-matter distribution (for early examples, see \cite{coles2, Fry:1992vr, McDonald:2009dh}, while for a modern review, see~\cite{Desjacques:2016bnm}). Within the EFT of LSS, dark matter and baryons are treated as effective fluids whose long-wavelength densities and momenta evolve according to coarse-grained equations of motion, while the galaxy density is represented as a composite operator depending on these long-wavelength fields. This perspective leads naturally to the bias expansion, a Taylor series in dark-matter fluctuations and their spatial derivatives, capturing how galaxy formation is affected by the local gravitational environment.   As discussed originally in \cite{Senatore:2014eva} and more recently and explicitly in \cite{Anastasiou:2025jsy} (see also \cite{Mirbabayi:2014zca, Baldauf:2014qfa, Angulo:2015eqa, Desjacques:2016bnm, DAmico:2022osl, DAmico:2022ukl, Donath:2023sav, Ansari:2024efj} for related works), the long-wavelength density of galaxies (and other long-wavelength quantities such as the dark-matter stress tensor) at the spacetime point $(\xvec , t)$ can in general depend on the full past light cone of~$(\xvec , t)$.  However, since most matter does not move very far during the history of the universe, the spatial dependence of clustering signals can be well represented by a spatial derivative expansion.   There is, however, no such hierarchy of scales for the time dependence: the time scale of the modes that are integrated out is about the same as the time scale of the full history of the universe.  This makes galaxy clustering, and the EFT of LSS in general, local in space, but non-local in time.

This extremely powerful approach has enabled robust analyses of galaxy clustering data, with impressive results.  One of the key advantages of the EFT approach is that it is based on basic principles, and is agnostic to specific details about obscure small-scale physics.  This allows one to controllably analyze \emph{all} the data, and \emph{down to smaller scales}, which in turn delivers more statistical power.  The first EFT of LSS analyses (sometimes called `full shape analyses') \cite{DAmico:2019fhj, Ivanov:2019pdj, Colas:2019ret} were done with the one-loop power spectrum (i.e. two-point function) on BOSS data \cite{BOSS:2016wmc}, with~\cite{DAmico:2019fhj} also including the tree-level bispectrum (i.e. three-point function).  Additional analyses that included the tree-level bispectrum are \cite{Philcox:2021kcw, Cabass:2022wjy, Cabass:2022ymb, Ivanov:2023qzb}, and the one-loop bispectrum was then analyzed in \cite{DAmico:2022osl, DAmico:2022gki}.  Extensions beyond the baseline analysis and use on other data sets include \cite{DAmico:2020kxu, DAmico:2020ods,  DAmico:2020tty, Zhang:2021yna,  Zhang:2021uyp, Simon:2022adh, Simon:2022csv, Piga:2022mge,  Chudaykin:2022nru, Glanville:2022xes, Spaar:2023his, Zhao:2023ebp, Cabass:2024wob, Lu:2025gki, DAmico:2025zui, Cagliari:2025rqe}.  The same techniques have now been adopted by the DESI collaboration in recent full-shape analyses \cite{DESI:2024jxi, DESI:2024hhd, Elbers:2025vlz}, and independently used on DESI data in \cite{Chudaykin:2025aux, Chudaykin:2025lww, Chudaykin:2025vdh}.

In this work, we focus on deriving the bias expansion for the overdensity of galaxies $\delta_g ( \xvec , t )$, which is given by
\be
\delta_g ( \xvec , t) = \frac{n_g ( \xvec , t ) - \bar n_g (t)}{\bar n_g(t)} \ , 
\ee
where $n_g ( \xvec , t ) $ is the number density of galaxies, and $\bar n_g(t)$ is the background number density of galaxies. 
The general expression for the galaxy overdensity at point $(\xvec , t)$ is \cite{Senatore:2014eva}
\begin{align}
\begin{split} \label{nggeneq}
\delta_g (\vec x,t) = N_g \left(\left\{H_0, \omn ,\dots,m_{\rm dm}, g_{\rm ew}, \ldots, \partial_i \partial_j \Phi,\partial_i v^j,\ldots\right\}_{{\rm past\, light\, cone\,} (\vx,t)}\right)  \ , 
\end{split}
\end{align}
where $N_g$ is some very complicated function (that presumably only Nature knows exactly),  $H_0$ is the Hubble constant, $\omn$ is the present dark matter abundance, $m_{\rm dm}$ is the mass of the dark matter, $g_{\rm ew}$ is the weak coupling constant, $\Phi$ is the gravitational potential, and $v^i$ is the dark-matter velocity.  Indeed, $N_g$ can depend on a large number of physical parameters, both known and unknown.  Notice that the equivalence principle means that only second spatial derivatives of the gravitational potential and gradients of the velocity can appear in \eqn{nggeneq}.\footnote{For the case of multiple large-scale fluids, such as including baryons \cite{Angulo:2015eqa}, the relative velocity can appear in \eqn{nggeneq} without the gradient.  In any case, these effects are expected to be small \cite{Lewandowski:2014rca, Braganca:2020nhv}, so what we will compute here is expected to be the leading contribution.  However, given that we are going to such a high perturbative order, it would be interesting to study the effect of baryons.  We leave this for future work.}  We use the subscript ${}_{{\rm past\, light\, cone\,} (\vec x,t)}$ to indicate that the variables are evaluated on the past light cone of the spacetime point $(\vx,t)$.  We can simplify this expression by recognizing that in a Hubble time, matter does not move much more than the non-linear scale, about $10$ Mpc.  Thus, the influence of the past light cone simplifies to the influence of the past tube around the fluid flow that ended up at $(\xvec , t)$, 
\begin{align}
\begin{split} \label{nggeneq2}
\delta_g (\vec x,t) = N_g \left(\left\{H_0, \omn ,\dots,m_{\rm dm}, g_{\rm ew}, \ldots, \partial_i \partial_j \Phi,\partial_i v^j,\ldots\right\}_{{\rm past\, tube\,} (\vx,t)}\right)  \ , 
\end{split}
\end{align}
where the subscript ${}_{{\rm past\, tube\,} (\vx,t)}$ indicates the past fluid tube with spatial size of about $10$ Mpc and temporal extent of about a Hubble time.

Given that we are interested in the galaxy density on large scales, where the matter fluctuations are small, we can Taylor expand \eqn{nggeneq2} with respect to these small fluctuations.  This means that we Taylor expand in the fields $r^{ij}$ and $p^{ij}$
\be \label{randpdefs}
r^{ij} \equiv \frac{2 \pd_i\pd_j \Phi}{3 \om a^2 H^2} \andd  p^{ij} \equiv -  \frac{\pd_i v^j }{a H f }  \ ,
\ee
which are the building blocks of the bias expansion.  Here, for convenience, we have normalized with some time-dependent factors.  $\Om \equiv \Om(a)$ is the time-dependent dark-matter fraction, $a(t)$ is the scale factor, $H \equiv H(a)$ is the Hubble rate, $D \equiv D(a)$ is the linear growth rate (see \eqn{daeqn}), and $f \equiv f(a)$ is the linear growth function (see above \eqn{edsdelta}).  To form the scalar $\delta_g$, we can contract the indices on $r^{ij}$ and $p^{ij}$ with the Kronecker delta $\delta^K_{ij}$.\footnote{This is for parity preserving physics, which we focus on in this work.  If parity is violated, one can contract the indices with the Levi-Civita tensor $\epsilon^{ijk}$.}   We will always denote the traces $\delta_{ij}^K r^{ij} = \delta$ (which is true because of the Poisson equation \eqn{poisson}) and $\delta_{ij}^K p^{ij} \equiv \theta$ (which is our definition of the rescaled velocity divergence $\theta$). For details on the dark-matter solution, see \appref{dmsolapp}.  To shorten notation, we use square brackets to denote traces in the given order, e.g. $[rprp] \equiv r^{ij} p^{jk}r^{kl}p^{li}$ (with sequential indices summed over), and powers inside of brackets signify matrix multiplication, as in $[r^2 p] \equiv r^{ij}r^{jk}p^{ki}$. Notice that while $r^{ij}$ is obviously symmetric, $p^{ij}$ is in general not.  However, if vorticity can be ignored (which for our purposes it can be, see \appref{vorticityapp}), then $p^{ij}$ is symmetric.

Since the general dependence of $\delta_g (\vec x,t)$ on $r^{ij}$ and $p^{ij}$ is non-local in time, this means that those fields evaluated at each past time should be treated as a separate variable when doing the Taylor expansion, and so we end up integrating over all possible times (i.e. summing over all variables), where each field can appear at a different time.  As explicitly detailed in \cite{Anastasiou:2025jsy}, we end up with an expression 
\begin{align}
\begin{split} \label{agenexpansion}
\delta_g ( \xvec , t ) = & \int^t d t_1 H(t_1 ) c_{\delta} ( t , t_1 ) \delta ( \xfl ( \xvec , t, t_1) , t_1) + \dots \\
& + \int^t d t_1 H(t_1 ) \int^t d t_2 H(t_2)  c_{[rp]} ( t , t_1, t_2 ) r^{ij} ( \xfl ( \xvec , t , t_1 ) , t_1 ) p^{ji} ( \xfl ( \xvec , t , t_2 ) , t_2 ) + \dots  \\
& +  \int^t d t_1 H(t_1 ) \int^t d t_2 H(t_2)   \int^t d t_3 H(t_3)  c_{[rpr]} ( t , t_1, t_2, t_3  ) \\
& \hspace{1in} \times  r^{ij} ( \xfl ( \xvec , t , t_1 ) , t_1 ) p^{jk} ( \xfl ( \xvec , t , t_2 ) , t_2 )  r^{ki} ( \xfl ( \xvec , t , t_3 ) , t_3 ) + \dots  \\
& + \dots  \ ,
\end{split}
\end{align}
where $\xfl$ is the fluid trajectory, see \eqn{xflintdef}, and the $c_{\{\delta , [rp], [rpr], \dots \}}$ are unknown time-dependent functions.  In the above expression, the ellipsis on each line indicates that we sum over all scalar contractions of $r^{ij}$ and $p^{ij}$ with the respective number of fields, and the final ellipsis indicates that we continue adding more and more powers of $r^{ij}$ and $p^{ij}$.  Since we are interested in the leading order in derivative bias expansion in this work, we have ignored the higher-derivative and stochastic terms above, which in any case can be easily included in this formalism, as explicitly outlined in \cite{Anastasiou:2025jsy}.  Other ways of constructing the bias expansion have been given in \cite{Mirbabayi:2014zca, Desjacques:2016bnm, Eggemeier:2018qae, Fujita:2020xtd, DAmico:2021rdb, Eggemeier:2021cam, Schmidt:2020ovm, Marinucci:2024add, Ansari:2025nsf}.

Our goal in this work is to formalize the perturbative expansion of \eqn{agenexpansion} for the galaxy overdensity (which is also applicable, for example, to the dark-matter stress tensor $\tau^{ij}$ with trivial modifications \cite{Anastasiou:2025jsy}) and use it to compute the bias expansion to sixth order.  Sixth order is relevant for computing the two-loop bispectrum and one-loop five-point function, for example,\footnote{Ambitiously, even the tree-level seven-point function.} but we leave a full analysis of these computations for future work.  As we will show, the non-local-in-time bias expansion at sixth order has 57 bias parameters, while the corresponding local-in-time expansion\footnote{This corresponds to the $c_{\{\delta , [rp], [rpr], \dots \}}$ functions in \eqn{agenexpansion} only having support at the final time $t$.} at sixth order has 46 bias parameters.  This means that at a fixed redshift, the galaxy clustering signal is sensitive to whether the galaxy and dark-matter halo distribution formed in a local-in-time way or in a non-local-in-time way.  Said another way, measurement of a sizable $\sim \mathcal{O}(1)$ value for one of the biases that must be zero for the local-in-time expansion would be a \emph{direct} indication of the formation time of galaxies.  This possibility was first explored in \cite{Donath:2023sav} where it was found that the first perturbative order in the galaxy density where one can tell the difference between local- and non-local-in-time structure formation is at fifth order.  

Along the way, we prove a set of recursion relations for the perturbative kernels resulting from expanding \eqn{agenexpansion}.  These relations, which relate kernels of different orders, are a direct result of the fact that we must integrate over the fluid trajectory in order to introduce time non-locality in a way that satisfies the equivalence principle.  The recursion relations also provide a streamlined way to compute all of the functions relevant for the bias expansion. Furthermore, we show that the kernels resulting from our construction at all orders satisfy specific multi-leg soft limits (i.e. when the sum of a subset of external momenta goes to zero) that were pointed out in \cite{DAmico:2021rdb}.  These soft limits are also a result of correctly introducing time non-locality, and are thus a consequence of the equivalence principle.   Interesting topics for future study, which should be straightforward applications of the methods explored in this paper, are the inclusion of higher-derivative bias including redshift-space distortions, baryonic feedback, velocity vorticity, stochastic fields, and the explicit computation of the two-loop bispectrum.

In \secref{galtranssec}, we discuss the symmetries of our system and present some properties of the fluid flow.  In \secref{mitnlsec}, we derive the kernels for the non-local-in-time expansion and a recursion relation that can be used to generate them.  In \secref{sixordersec} we give the sixth-order basis of descendants, which has 57 biases, and compare it to the local-in-time expansion, which has 46 biases.  Also in \secref{sixordersec}, we derive multi-leg soft limits for the kernels given in this paper.  Background material  and many computational details are given in the appendices.  

In the rest of this work, use the following notation for integration
\be
\int_{\kvec_1 , \dots , \kvec_n} \equiv \int \frac{d^3 k_1 }{(2 \pi)^3} \cdots \frac{d^3 k_n}{(2 \pi)^3 } \ ,  \quad \int_{\kvec_1 , \dots , \kvec_n}^{\kvec} \equiv \int_{\kvec_1 , \dots , \kvec_n} ( 2 \pi)^3 \delta_D ( \kvec - \sum_{i = 1}^n \kvec_i )  \ , 
\ee
where $\delta_D$ is the Dirac delta function, and our Fourier conventions are
\be
f ( \xvec , t ) = \int_{\kvec} f (\kvec ,  t ) \,  e^{i \kvec \cdot \xvec}  \ . 
\ee
For a three-dimensional vector $\kvec$, we write $k \equiv | \kvec |$ for the magnitude, and $\hat k \equiv \kvec / k$ for the unit vector parallel to $\kvec$.  For the sum of labeled vectors, we will often use the notation
\be
\pvec_{m;n} \equiv \pvec_m + \pvec_{m+1} + \dots + \pvec_{n-1} +  \pvec_{n} \ . 
\ee
We use Latin letters like $i,j,k,l$ to denote spatial indices, in general we do not distinguish between upper and lower spatial indices, and repeated indices imply summation.   We also use the prime to denote derivatives with respect to the scale factor, $g' \equiv \partial g / \partial a$, and the dot to denote derivatives with respect to cosmic time, $\dot g \equiv \partial g / \partial t$.  

%
%
\section{Diffeomorphisms and the fluid trajectory} \label{galtranssec}

{Ultimately, the equations governing our system are determined by General Relativity, and as such, they are covariant under diffeomorphisms.  For the non-relativistic limit in the Newtonian approximation (which is what is relevant for galaxy clustering), a useful gauge is the so-called Newtonian gauge, where the scalar part of the metric is}\footnote{If anisotropic stress is zero, we have $\Phi = \Psi$ from the Einstein equations.  Even though anisotropic stress can be generated in the EFT of LSS, it is a relativistic correction. }
\be
ds^2 = - ( 1 + 2 \Phi) dt^2 +a(t)^2 ( 1 - 2 \Psi) d\xvec^2 \ .
\ee
Once we are in the Newtonian gauge, there is a subset of diffeomorphisms that, up to relativistic corrections, keep us in the Newtonian gauge.  These are the so-called residual large diffeomorphisms \cite{Creminelli:2012ed, Hinterbichler:2012nm, Hinterbichler:2013dpa, Creminelli:2013mca, Horn:2014rta} 
\begin{align}
\begin{split}  \label{transf2}
& \tilde x^i = x^i +   n^i(t)   \ , \quad \tilde t  = t + a(t)^2 \nvec (t) \cdot \xvec  \ , \quad  \tilde v^i ( \tilde x^j , t )  = v^i( x^j , t) + a(t)  \dot n^i(t)  \ , \\
 &\tilde \Phi ( \tilde x^i , t )   = \Phi ( x^i , t) - a(t)^2 (\ddot n^i ( t) + 2 H(t) \dot n^i(t)) x^i  \ , \quad  \tilde \delta ( \tilde x^j , t ) = \delta ( x^j , t ) \ , 
\end{split}
\end{align}
for a generic time-dependent $\nvec (t)$.  The above transformations for $\delta$, $\vec{v}$, and $\Phi$ can be straightforwardly derived from the usual diffeomorphism transformation rules with respect to the change of coordinates in \eqn{transf2}.   Since the transformation \eqn{transf2} keeps us in the Newtonian gauge, the equations of motion are invariant under \eqn{transf2}.  

We point out that \eqn{transf2} is not a Galilean transformation  \cite{Creminelli:2013mca} (which would have $\nvec ( t) = \uvec \, t$ for a constant velocity $\uvec$).  Galilean transformations are indeed a symmetry of our system, but \eqn{transf2} is a consequence of diffeomorphism invariance, and so is inherited from General Relativity.   For brevity, we refer to \eqn{transf2} as an NRNGP (non-relativistic Newtonian-gauge preserving) transformation.  Already just looking at the particular subset of diffeomorphisms given by \eqn{transf2} shows that  that clustering can only depend on the velocity field with at least one spatial derivative, and the gravitational potential with at least two spatial derivatives (which of course one can deduce by looking at the full diffeomorphisms).  For the consequences of the NRNGP transformation to the case of two fluids, such as dark matter and baryons, see \cite{Braganca:2020nhv}. 

We can use \eqn{transf2} to define the transformation of a generic NRNGP scalar $\phi$ as
\be \label{scalartransfrule}
\tilde \phi ( \xvec + \nvec(t) , t) = \phi ( \xvec , t )  \ .
\ee
In particular, we see that $r^{ij} \propto \partial_i \partial_j \Phi$ and $p^{ij} \propto \partial_i v^j$ are both scalars under \eqn{transf2}.  Note that the coordinate change in $t$ is almost always a relativistic correction, but is used to derive the transformation of $\Phi$, for example (which comes from the normal transformation of $g_{\mu \nu}$ under diffeomorphisms).  Therefore, when it is irrelevant for the non-relativistic, Newtonian limit, we will leave off the transformation of the time coordinate, as we did in \eqn{scalartransfrule}.  

\begin{figure}[t!]
  \centering
  \includegraphics[height=0.42\textwidth]{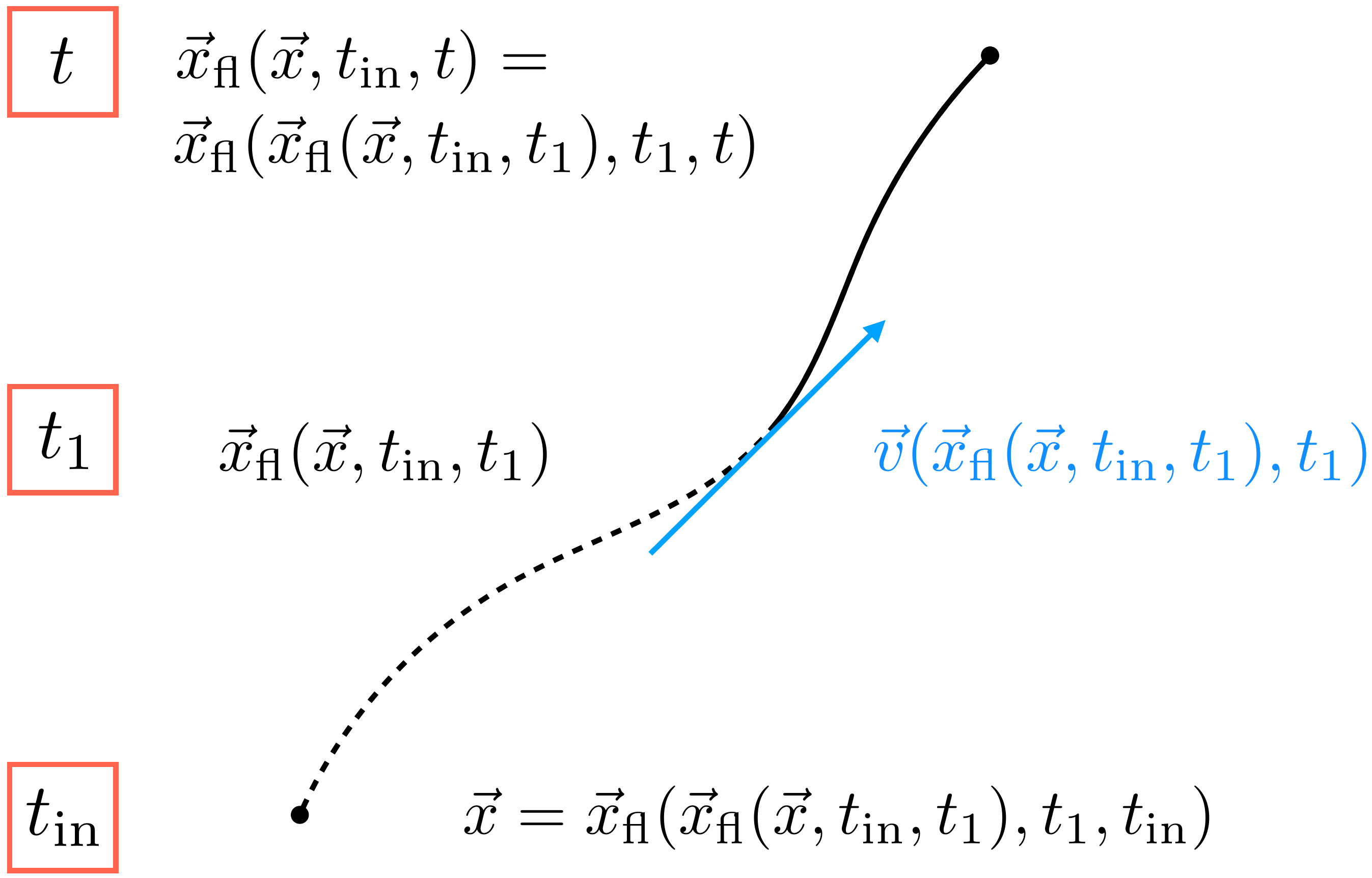}
  \caption{Diagrammatic representation of the composition rule for the fluid trajectory.}\label{xflfig}
\end{figure}

In this work, we are building the non-local-in-time bias expansion, so our main concern is that our non-local-in-time expression for $\delta_g ( \xvec , t)$ is a scalar under \eqn{transf2}.  To ensure this, a crucial object is the fluid trajectory which is given by
\be \label{xflintdef}
\xvec_{\rm fl} ( \xvec , t_{\rm in}, t) = \xvec + \int_{t_{\rm in}}^t \frac{ d t'}{a(t') }  \vec{v} \left( \xvec_{\rm fl} ( \xvec , t_{\rm in} , t' ) , t' \right) \ , 
\ee
and satisfies 
\be \label{xfldiffeq}
\frac{d}{dt }  \xvec_{\rm fl} ( \xvec , t_{\rm in} , t)  =  \frac{1}{a(t)}  \vec{v}  ( \xfl( \xvec , t_{\rm in} , t), t) \ .
\ee
Physically, the expression $\xvec_{\rm fl} ( \xvec , t_{\rm in}, t)$ means that one starts at position $\xvec$ at time $t_{\rm in}$, and then one follows the velocity field $\vec{v}$ up to time $t$, see \figref{xflfig}.

The fluid trajectory satisfies an intuitive composition rule
\be \label{comprule1}
\xfl \left( \xfl ( \xvec , t_{\rm in} , t_1 ) , t_1 , t \right) = \xfl ( \xvec , t_{\rm in} , t ) \ .
\ee
\eqn{comprule1} is simply the statement that if one starts at point $\xvec$ at time $\tin$, follows the fluid up to time $t_1$, and then follows the fluid from time $t_1$ to time $t$, this is the same as simply following the fluid from point $\xvec$ at $\tin$ directly to $t$.
In particular, by plugging $t = t_{\rm in}$ in \eqn{comprule1}, we also have the inverse relationship
\be \label{comprule2}
\xfl \left(  \xfl ( \xvec , t_{\rm in} , t_1 ) , t_1 , t_{\rm in} \right) = \xvec \ ,
\ee
since $\xfl ( \xvec , t_{\rm in} , \tin ) = \xvec$, as can be seen in \eqn{xflintdef}.  These relationships can be visualized in \figref{xflfig}.   
To rigorously prove these rules, we first look at \eqn{comprule1}, since \eqn{comprule2} follows from that.  We will show that \eqn{comprule1} holds by showing that, as a function of $t$, both the left- and right-hand sides pass through the same point and satisfy the same ordinary differential equation.  To show that they pass through the same point, evaluate both sides of \eqn{comprule1} at $t = t_1$.   The left-hand side is $\xfl \left( \xfl ( \xvec , t_{\rm in} , t_1 ) , t_1 , t_1 \right) $ which is equal to $\xfl ( \xvec , t_{\rm in} , t_1 ) $ since the integral from $t_1$ to $t_1$ in the definition \eqn{xflintdef} is zero.  The right-hand side is trivially equal to $\xfl ( \xvec , t_{\rm in} , t_1 ) $, so indeed both sides pass through the same spatial point at time $t = t_1$.  Now we show that they satisfy the same first-order differential equation.  The differential equation for the right-hand side of \eqn{comprule1} is just \eqn{xfldiffeq}.  To find the differential equation for the left-hand side of \eqn{comprule1}, we first replace $\tin$ with $t_1$ in \eqn{xfldiffeq}, and then we plug in $\xfl ( \xvec , \tin , t_1)$ for $\xvec$ in \eqn{xfldiffeq}.  This gives
\be
\frac{d}{dt }  \xvec_{\rm fl} ( \xfl ( \xvec , \tin , t_1) , t_{1} , t)  =\frac{1}{a(t)}  \vec{v}  ( \xfl( \xfl ( \xvec , \tin , t_1) , t_{1} , t), t) \ ,
\ee
which indeed is the same equation for  $\xfl \left( \xfl ( \xvec , t_{\rm in} , t_1 ) , t_1 , t \right) $ as \eqn{xfldiffeq} is for $\xfl ( \xvec , t_{\rm in} , t )$.  Therefore, by the uniqueness of solutions to first-order differential equations, the two must be the same functions, and \eqn{comprule1} holds.  

Another useful property is the derivative of the fluid trajectory with respect to the starting time.  Let us start with \eqn{comprule1} in the form 
\be \label{comprule3}
\xfl \left( \xfl ( \xvec , t_{\rm in} , t ) , t , t '  \right) = \xfl ( \xvec , t_{\rm in} , t '  ) \ .
\ee
Since the right-hand side is independent of the intermediate point $t$, we have
\be
\frac{d}{dt} \xfl \left( \xfl ( \xvec , t_{\rm in} , t ) , t , t '  \right)  = 0 \ ,
\ee
which gives
\be
\left[ \frac{\partial}{\partial t } \xfl ( \yvec , t , t' ) + \frac{d}{dt} x_{\rm fl}^j ( \xvec , \tin , t) \frac{\partial}{\partial y^j} \xfl ( \yvec , t , t' )  \right]\Big|_{\yvec = \xfl ( \xvec , \tin , t )} = 0 \ , 
\ee
and then using \eqn{xfldiffeq}, this gives 
\be
\left[ \frac{\partial}{\partial t } \xfl ( \yvec , t , t' ) + \frac{1}{a(t)} v^j ( \yvec , t ) \frac{\partial}{ \partial y^j} \xfl ( \yvec , t , t' )  \right]\Big|_{\yvec = \xfl ( \xvec , \tin , t )} = 0 \ .
\ee
Now, since we can take the initial time to be whatever we want, we can take $\tin = t$, and then we simply have
\be \label{otherxflderiv}
\frac{\partial}{\partial t } \xfl ( \xvec , t , t' ) + \frac{1}{a(t)} v^j ( \xvec , t ) \frac{\partial}{\partial x^j} \xfl ( \xvec , t , t' ) = 0 \ , 
\ee
for all $\xvec$.  

We can use the above results to prove some useful identities that we will use later.  For generic functions $f( \xvec , t)$, \eqn{xfldiffeq} implies
\be
\frac{d}{dt} f ( \xfl ( \xvec , t' , t ) , t)  = \left[ \frac{\partial}{\partial t} f ( \yvec , t ) + \frac{1}{a(t)} v^i ( \yvec , t) \frac{\partial}{\partial y^i} f ( \yvec , t) \right] \Big|_{\yvec = \xfl ( \xvec , t' , t) }  \ ,
\ee
and \eqn{otherxflderiv} implies 
\begin{align}
\begin{split} \label{timederivxflf2}
 \frac{d}{dt ' } f ( \xfl ( \xvec , t' , t ) , t) &  = \frac{d \, x^j_{\rm fl}(\xvec , t' , t) }{dt'} \frac{\partial f ( \yvec , t)}{\partial y^j} \Bigg|_{\yvec = \xfl ( \xvec , t' , t )}  \\
 & = -  \frac{1}{a(t')} v^i ( \xvec , t ') \frac{\partial}{\partial x^i} x^j_{\rm fl} ( \xvec , t' , t ) \frac{\partial f ( \yvec , t)}{\partial y^j} \Bigg|_{\yvec = \xfl ( \xvec , t' , t )}  \\
& =  - \frac{v^i ( \xvec , t') }{a(t')} \frac{\partial}{\partial x^i} f ( \xfl ( \xvec , t' , t) , t)  \ ,
\end{split}
\end{align}
where in the last line we used the chain rule.

\begin{figure}[t!]
  \centering
  \includegraphics[height=0.45\textwidth]{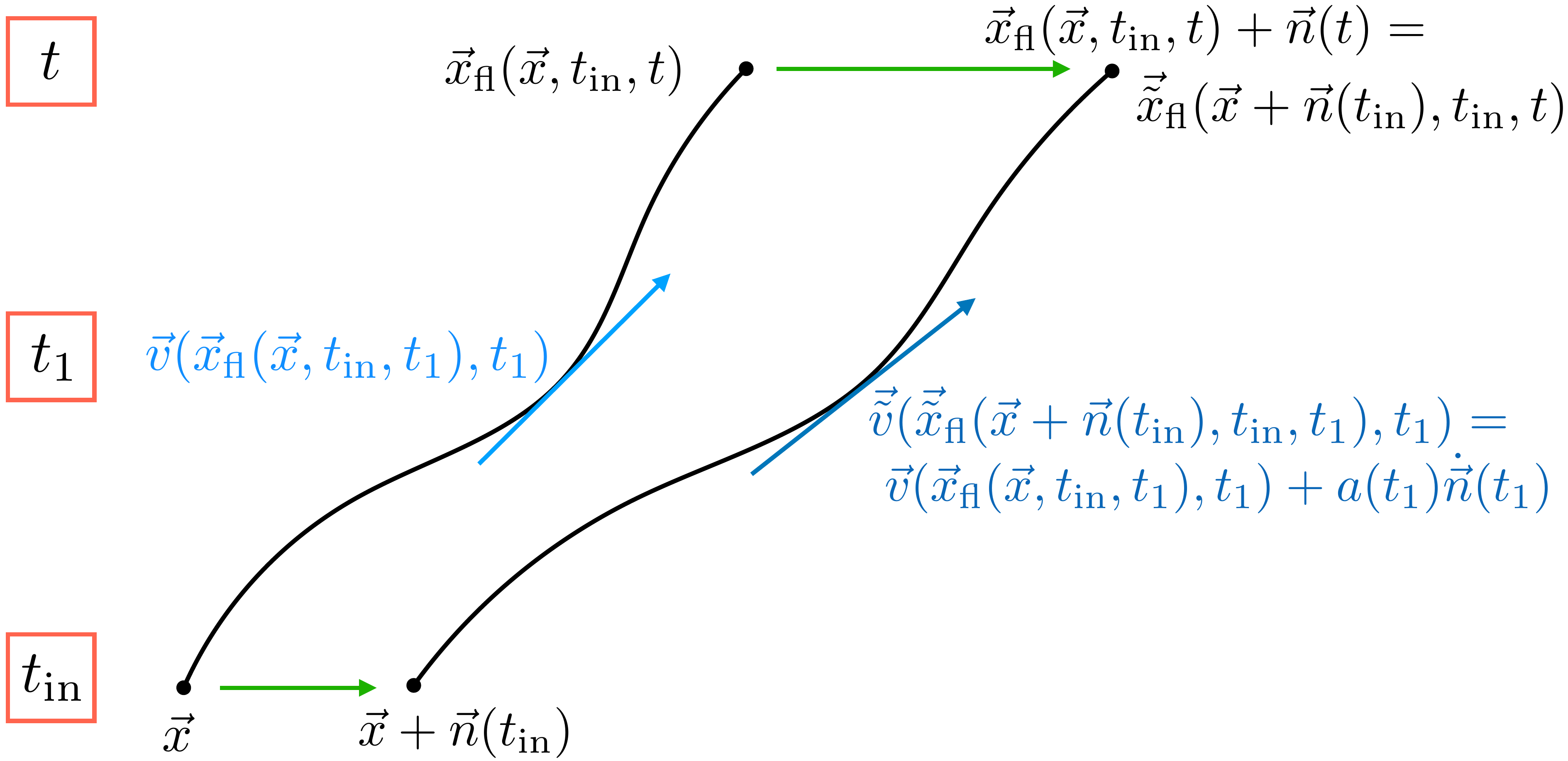}
  \caption{Diagrammatic representation of the NRNGP transformation for the fluid trajectory.}\label{xflfig2}
\end{figure}

We can now derive how the fluid trajectory transforms under an NRNGP transformation.  After an NRNGP transformation, the new fluid trajectory $\vec{\tilde{x}}_{\rm fl}$ should be the path traced by following the new velocity $\vec{\tilde{v}}$, i.e., 
\be \label{xfltildediffeq}
\frac{d}{dt }  \vec{\tilde{x}}_{\rm fl} ( \xvec , t_{\rm in} , t)  =\frac{1}{a(t)}  \vec{\tilde{v}}  ( \vec{\tilde{x}}_{\rm fl} ( \xvec , t_{\rm in} , t), t) \ .
\ee
Expressing this equation in terms of the old fields must be equivalent to the original equation \eqn{xfldiffeq}.  To start our manipulations, let us first evaluate \eqn{xfltildediffeq} at $\xvec \rightarrow \xvec + \nvec(t_{\rm in})$ 
\be \label{xfltildediffeqp}
\frac{d}{dt }  \vec{\tilde{x}}_{\rm fl} ( \xvec  + \nvec(\tin) , t_{\rm in} , t)  =\frac{1}{a(t)}  \vec{\tilde{v}}  ( \vec{\tilde{x}}_{\rm fl} ( \xvec + \nvec(\tin)  , t_{\rm in} , t), t) \ ,
\ee
and then use the transformation law \eqn{transf2} for $\vvec$ to get
\be \label{xfltildediffeqpp}
\frac{d}{dt }  \vec{\tilde{x}}_{\rm fl} ( \xvec  + \nvec(\tin) , t_{\rm in} , t)  =\frac{1}{a(t)}  \left( \vec{v}  ( \vec{\tilde{x}}_{\rm fl} ( \xvec + \nvec(\tin)  , t_{\rm in} , t) - \nvec(t), t)   + a(t) \dot{\nvec}(t) \right) \ .
\ee
We want to solve for $ \vec{\tilde{x}}_{\rm fl} ( \xvec  + \nvec(\tin) , t_{\rm in} , t)$ in terms of $\xfl ( \xvec , t_{\rm in} , t )$ such that \eqn{xfltildediffeqpp} is equivalent to \eqn{xfldiffeq}. The unique solution that satisfies $\vec{\tilde{x}}_{\rm fl} ( \xvec   , t_{\rm in} , \tin ) = \xvec $ (which is a desirable property that we impose by choosing the integration constant for the solution of \eqn{xfltildediffeq}) is 
\be \label{xfltransform}
\vec{\tilde{x}}_{\rm fl} ( \xvec + \nvec(\tin)  , t_{\rm in} , t ) = \xfl ( \xvec , t_{\rm in} , t ) + \nvec(t) \ , 
\ee
which gives us the transformation law for $\xfl$.   To see this, we first show that \eqn{xfltransform} is indeed a solution.  First, plug \eqn{xfltransform} into \eqn{xfltildediffeqpp} to replace $\vec{\tilde{x}}_{\rm fl}$ for $\xfl$ to get
\begin{align}
\begin{split} \label{xflcheck}
\frac{d}{dt } \left[ \xfl ( \xvec , t_{\rm in} , t ) + \nvec(t) \right] &  =  \frac{1}{a(t)}  \left( \vec{v}  (\left[ \xfl ( \xvec , t_{\rm in} , t ) + \nvec(t) \right] - \nvec(t), t)   + a(t) \dot{\nvec}(t) \right)  \\
& =  \frac{1}{a(t)}  \vec{v}  ( \xfl( \xvec , t_{\rm in} , t), t) + \dot{\nvec}(t) \ ,
\end{split}
\end{align}
and then cancel the $\dot{\nvec}(t)$ on both sides to recover \eqn{xfldiffeq}.  Now, to show it is the only solution, we modify the right-hand side of \eqn{xfltransform} by adding an arbitrary function $\vec{g}(\xvec , \tin , t)$ which we will show must be zero. First of all, in order to have $\vec{\tilde{x}}_{\rm fl} ( \xvec   , t_{\rm in} , \tin ) = \xvec $, we must have $\vec{g}(\xvec , \tin , \tin ) = 0$.  Then, following manipulations similar to \eqn{xflcheck}, we find that the equation for $\vec{g}$ is
\be \label{gdiffeq}
\frac{d}{dt}\vec{g}(\xvec , \tin, t) = \frac{1}{a(t)} \left( \vec{v}  ( \xfl( \xvec , t_{\rm in} , t) + \vec{g}( \xvec , \tin, t) , t) - \vec{v}  ( \xfl( \xvec , t_{\rm in} , t), t)  \right)  \ ,
\ee
and since $\vec{g}(\xvec , \tin , \tin ) = 0$, \eqn{gdiffeq} implies that $\dot{\vec{g}}(\xvec , \tin , \tin ) = 0$, so we must have $\vec{g}(\xvec , \tin, t) = 0$ for all $t$.  The physical interpretation of \eqn{xfltransform} is such that we end up at the same place whether we start at $\xvec$, move forward with the original velocity, and then shift by $\nvec(t)$ at the very end, or we start with the shift $\nvec(\tin)$ at the beginning and move forward with the shifted velocity. This is shown pictorially in \figref{xflfig2}. 

We can also verify that our solution \eqn{xfltransform} is consistent at the level of the integral expressions (which of course it must be).  By integrating \eqn{xfltildediffeq}, we have
\begin{align} 
\vec{\tilde{x}}_{\rm fl} ( \xvec + \nvec(\tin) , t_{\rm in}, t) & = \xvec + \nvec(\tin) + \int_{t_{\rm in}}^t \frac{ d t'}{a(t') }  \vec{\tilde{v}} \left( \vec{\tilde{x}}_{\rm fl} ( \xvec + \nvec(\tin) , t_{\rm in} , t' ) , t' \right) \nonumber \\
& = \xvec + \nvec(\tin) + \int_{t_{\rm in}}^t \frac{ d t'}{a(t') } \left(   \vec{v} \left( \vec{\tilde{x}}_{\rm fl} ( \xvec + \nvec(\tin) , t_{\rm in} , t' )  - \nvec(t') , t' \right) + a(t') \dot{\nvec}(t')  \right) \nonumber  \\
& = \xvec +  \nvec(t)  + \int_{t_{\rm in}}^t \frac{ d t'}{a(t') }  \vec{v} \left( \vec{x}_{\rm fl} ( \xvec , t_{\rm in} , t' )   , t' \right)  \label{xfltildeintdef2} \\ 
& = \xfl ( \xvec , t_{\rm in} , t ) + \nvec(t) \ ,  \nonumber
\end{align}
where in the second line we used the transformation law \eqn{transf2} for $\vvec$, and in the third line we used the transformation \eqn{xfltransform} of $\xfl$ on the right-hand side.  In the last line we see that we end up with the same expression \eqn{xfltransform}, and so it is indeed consistent.   \eqn{xfltransform} was previously derived in \cite{Baldauf:2014qfa}.  

Using the above NRNGP transformation properties, we can also derive how NRNGP scalars transform when evaluated along the fluid trajectory.  For an NRNGP scalar $\phi ( \xvec , t)$, consider the composition $\phi ( \xvec_{\rm fl} ( \xvec , t , t' ) , t ' ) $.  We can express this in terms of the tilde fields by
\begin{align}
\begin{split} \label{galscalcomp}
\phi ( \xvec_{\rm fl} ( \xvec , t , t' ) , t ' )  & = \tilde \phi ( \xvec_{\rm fl} ( \xvec , t , t' ) + \nvec(t') , t ' )  = \tilde \phi ( \vec{\tilde{x}}_{\rm fl} ( \xvec + \nvec(t) , t , t' )  , t ' ) \ . 
\end{split}
\end{align}
This means that the composition 
\be
\mathcal{O} ( \xvec , t , t') \equiv \phi ( \xvec_{\rm fl} ( \xvec , t , t' ) , t ' )  \ , 
\ee
transforms as an NRNGP scalar with coordinates $( \xvec, t)$, where
\be
\tilde{\mathcal{O}} ( \xvec , t , t') \equiv \tilde \phi ( \vec{\tilde{x}}_{\rm fl} ( \xvec , t , t' ) , t ' )  \ ,
\ee
i.e.
\be
\tilde{\mathcal{O}}( \xvec + \nvec ( t) , t , t') = \mathcal{O} ( \xvec , t , t') \ ,
\ee
which is the transformation of an NRNGP scalar \eqn{scalartransfrule}.

%
%
\section{Non-local-in-time structure formation} \label{mitnlsec}

We now give a general constructive method to compute all non-local-in-time kernels for the galaxy bias.  First we specialize the expansion of \cite{Senatore:2014eva, Anastasiou:2025jsy} to the case of scalars (such as the galaxy overdensity), and then we will derive a simple recursion relation to generate all of the relevant functions.  The goal of this section is to find all of the potentially independent functional forms that can appear; later, we will explicitly solve for degeneracies of these functions at sixth order to find the minimal non-local-in-time basis (i.e. bias expansion).  We note that the general construction given in this section does not rely on $p^{ij}$ being symmetric (i.e. it is also applicable to the inclusion of vorticity).

%
%
\subsection{Perturbative expansion}

First, using the results of \secref{galtranssec}, it is easy to show that our construction in \eqn{agenexpansion} is indeed an NRNGP scalar.  As an example, consider the term
\be
\mathcal{O}_2 ( \xvec , t ) =  \int^t d t_1 H(t_1 ) \int^t d t_2 H(t_2)  c_{[rp]} ( t , t_1, t_2 ) r^{ij} ( \xfl ( \xvec , t , t_1 ) , t_1 ) p^{ij} ( \xfl ( \xvec , t , t_2 ) , t_2 )  \ ,
\ee
for which we have the transformed field 
\be
\tilde{\mathcal{O}}_2 ( \xvec , t ) =  \int^t d t_1 H(t_1 ) \int^t d t_2 H(t_2)  c_{[rp]} ( t , t_1, t_2 ) \tilde r^{ij} ( \vec{\tilde{x}}_{\rm fl} ( \xvec , t , t_1 ) , t_1 ) \tilde p^{ij} ( \vec{\tilde{x}}_{\rm fl} ( \xvec , t , t_2 ) , t_2 )  \ .
\ee
Plugging in for $\xvec$ the value $ \xvec + \nvec(t)$, we have 
\begin{align}
\tilde{\mathcal{O}}_2 ( \xvec + \nvec(t), t ) & =   \int^t d t_1 H(t_1 ) \int^t d t_2 H(t_2)  c_{[rp]} ( t , t_1, t_2 )\\
&\quad \quad \times \tilde r^{ij} ( \vec{\tilde{x}}_{\rm fl} ( \xvec + \nvec(t) , t , t_1 ) , t_1 ) \tilde p^{ij} ( \vec{\tilde{x}}_{\rm fl} ( \xvec  + \nvec(t) , t , t_2 ) , t_2 ) \nonumber \\
& = \mathcal{O}_2 ( \xvec , t)  \ ,
\end{align}
where we used \eqn{galscalcomp} to replace $\tilde r^{ij}$ and $\tilde p^{ij}$ in terms of $r^{ij}$ and $p^{ij}$.  This is exactly the transformation rule for NRNGP scalars.  This logic obviously generalizes to any number of factors of $r^{ij}$ and $p^{ij}$ since each field can be replaced separately, and so it is clear that the full expression \eqn{agenexpansion} is an NRNGP scalar.  Now, for convenience, let us rewrite \eqn{agenexpansion} as\footnote{Here and elsewhere, we use the shorthand notation
\be
\int^t \, d t_1 \cdots d t_m \equiv \int^t \, d t_1 \cdots  \int^t  d t_m  \ , 
\ee
to remove clutter. }
\begin{align}
\begin{split} \label{generaltauij}
\delta_g^{(n)} ( \xvec , t ) & = \sum_{m = 1}^n \sum_{\{\mathcal{O}_m\}} \int^t \, d t_1 \cdots d t_m\, H ( t_1 ) \cdots H(t_m )  \, c_{\mathcal{O}_m } ( t , t_1 , \dots , t_m ) \\
 & \hspace{1in} \times  \left[ \mathcal{O}_{m} \big|_{\xfl}  ( \xvec , t ; t_1 , \dots , t_m )  \right]^{(n)}  \ .
\end{split}
\end{align}
The set $\{ \mathcal{O}_m \}$ is the set of all scalars built out of $r^{ij}$ and $p^{ij}$ that have $m$ factors, and the notation $\mathcal{O}_{m} \big|_{\xfl}$ represents the fluid expansion at multiple times, which we are going to explain below.

Since each $\mathcal{O}_m$ is the product of $m$ factors of $r^{ij}$ and $p^{ij}$, we can write each field explicitly as 
\be \label{omfactors}
\mathcal{O}_m ( \xvec , t ) = t^{i_1 j_1}_{\mathcal{O}_m , 1} ( \xvec , t )  \cdots t^{i_m j_m}_{\mathcal{O}_m , m} ( \xvec , t ) \, \Delta_{\mathcal{O}_m} [ i_1 , j_1 , \dots , i_m , j_m ] \ ,
\ee
where each $t^{i_k j_k}_{\mathcal{O}_m , k}$ is either $r^{i_k j_k}$ or $p^{i_k j_k }$, and $\Delta_{\mathcal{O}_m} [ i_1 , j_1 , \dots , i_m , j_m ] $ is the combination of Kronecker delta functions needed to build $\mathcal{O}_m$.  For example, for $\mathcal{O}_3 = r^{i j } p^{jk} r^{ki}$, we have $t_{\mathcal{O}_3, 1}^{i_1 j_1} = r^{i_1 j_1}$, $t_{\mathcal{O}_3, 2}^{i_2 j_2} = p^{i_2 j_2}$, $t_{\mathcal{O}_3, 3}^{i_3 j_3} = r^{i_3 j_3}$, and  $\Delta_{\mathcal{O}_3} [ i_1 , j_1 , i_2 , j_2, i_3, j_3 ] = \delta_K^{ i_1 j_3} \delta_K^{j_1 i_2} \delta_K^{j_2 i_3}$.    Given this, we use the following notation for the fluid expansion at multiple times, where each factor is evaluated at a different time on the fluid trajectory, 
\begin{align}
\begin{split} \label{oxflprodexp}
& \mathcal{O}_{m} \big|_{\xfl}  ( \xvec , t ; t_1 , \dots , t_m ) \equiv t^{i_1 j_1}_{\mathcal{O}_m , 1} ( \xfl ( \xvec , t , t_1) , t_1 )  \cdots t^{i_m j_m}_{\mathcal{O}_m , m} ( \xfl ( \xvec , t , t_m)  , t_m ) \\
&\hspace{2in}  \times  \Delta_{\mathcal{O}_m} [ i_1 , j_1 , \dots , i_m , j_m ] \ .
\end{split}
\end{align}
To find $\left[ \mathcal{O}_{m} \big|_{\xfl}  ( \xvec , t ; t_1 , \dots , t_m )  \right]^{(n)} $, we then Taylor expand the fluid trajectory $\xfl$ and take the $n$-th order piece.  This can be done in a number of ways.  One way is to iteratively expand the definition \eqn{xflintdef} in powers of $v^i$, and then expand the spatial argument in each $t^{i_k j_k}_{\mathcal{O}_m , k} ( \xfl ( \xvec , t , t_k) , t_k )$.  A simpler way is to use the recursion relation of \cite{Donath:2023sav}, which we detail in \appref{pertexpapp}.\footnote{At the beginning of \secref{sixordersec}, we explain the difference of our procedure with respect to the one of \cite{Donath:2023sav}.}

The final result is that we can express
\begin{align}
\begin{split}  \label{finalfulltauij}
\delta_g^{(n)} ( \xvec , t ) & = \sum_{m = 1}^n \sum_{\{\mathcal{O}_m\}} \sum_{\{\alpha_1, \dots , \alpha_m\}_{\leq n} } c_{\mathcal{O}_m; \alpha_1 , \dots , \alpha_m } (t )  \bar{ \mathbb{C}}^{(n)}_{\mathcal{O}_m; \alpha_1 , \dots , \alpha_m} ( \xvec , t) \ ,
\end{split}
\end{align}
where $\{\alpha_1, \dots , \alpha_m\}_{\leq n} $ is the set of all $(\alpha_1 , \dots , \alpha_m)$ with $\alpha_i \geq 1$ and  $\alpha_1 + \cdots + \alpha_m \leq n$.  The kernels $\bar{ \mathbb{C}}^{(n)}_{\mathcal{O}_m; \alpha_1 , \dots , \alpha_m} $ are given by 
\be \label{finalcndef}
\bar{\mathbb{C}}^{(n)}_{\mathcal{O}_m; \alpha_1 , \dots , \alpha_m}  ( \xvec , t) =  \sum_{\{n_1 , \dots , n_m\}^\alpha_n }     \Delta_{\mathcal{O}_m} [ i_1 , j_1 , \dots , i_m , j_m ]   \prod_{a = 1}^m   [ \mathbb{C}_{ (\mathcal{O}_m , a)  , \alpha_a}^{ i_a j_a}]^{(n_a)} ( \xvec , t )    \ , 
\ee
where $\{n_1 , \dots , n_m\}^\alpha_n $ is the set of all $(n_1 , \dots , n_m )$ such that $n_1 + \cdots + n_m = n$ and $n_a \geq \alpha_a$ for $a = 1 , \dots , m$, and the kernels $ \mathbb{C}_{ (\mathcal{O}_m , a)  , \alpha_a}^{ i_a j_a}$ can be obtained following \cite{Donath:2023sav}, and are also given in \appref{pertexpapp}.  Notice that the bar $\bar{\mathbb{C}}$ indicates the bias kernels, while the kernels with no bar $\mathbb{C}$ are the building blocks and can be easily derived by the recursion \eqn{singlerec} and \eqn{singleet}.  This means that the initial set of potentially independent functions (i.e. before we check for degeneracies) is one for each $\mathcal{O}_m$ and each tuple $(\alpha_1 , \dots , \alpha_m)$ with $\alpha_1 + \cdots + \alpha_m \leq n$ and $\alpha_i \geq 1$.  

Thus, we have been lead to a straightforward way to compute all of the kernels.  First, compute all of the $[\mathbb{C}^{ij}_{r,\alpha}]^{(n)}$ and $[\mathbb{C}^{ij}_{p,\alpha}]^{(n)}$ using the recursion relations of \cite{Donath:2023sav} to the desired order $n$ (i.e. using \eqn{singlerec} and \eqn{singleet}).  Then, at $n$-th order, form all the sets $\{\mathcal{O}_m\}$ for $m \leq n$, where $\{\mathcal{O}_m\}$ is the set of all distinct trace products of $r^{ij}$ and $p^{ij}$ that contains $m$ factors, i.e.
\be
\{ \mathcal{O}_2 \} = \{ \theta^2 , \theta \delta , \delta^2, [p^2], [pr], [r^2] \} \ .
\ee
Then, for each $\mathcal{O}_m \in \{ \mathcal{O}_m \}$, find the set $\{ \alpha_1 , \dots , \alpha_m \}_{\leq n}$ of all tuples $(\alpha_1 , \dots , \alpha_m)$ with $\alpha_1 + \dots + \alpha_m~\leq~n$ and $\alpha_i \geq 1$.  Here, the ordering of the $\alpha_i$ in $(\alpha_1 , \dots , \alpha_m)$ matters because of the ordering of the products of $r^{ij}$ and $p^{ij}$ in $\mathcal{O}_m$.  Each one of these tuples for each $\{\mathcal{O}_m\}$ for $1 \leq m \leq n$ represents a function (given explicitly by \eqn{finalcndef}) to add to the list of possibly independent functions.

%
%
\subsection{Recursion relation}

Similar to the case presented in \cite{Donath:2023sav}, we can further simplify the construction of the kernels $\bar{\mathbb{C}}^{(n)}_{\mathcal{O}_m; \alpha_1 , \dots , \alpha_m} $ by deriving relationships between kernels of different orders.  Specifically, we will find a recursion relation relating kernels at order $n$ with   $\sum_{i =1 }^m \alpha_i~<~n$ to lower-order kernels, and then we give a simple expression for the remaining kernels with $\sum_{i =1 }^m \alpha_i~=~n$.

Equating \eqn{generaltauij} and \eqn{finalfulltauij}, and plugging the definition of the $c_{\mathcal{O}_m; \alpha_1, \dots , \alpha_m}$ \eqn{gentimefns} into \eqn{finalfulltauij}, we have\footnote{ To see this, set the right-hand sides of  \eqn{generaltauij} and \eqn{finalfulltauij} equal to each other (since they are just two equivalent ways to write $\delta_g^{(n)}$) to get
\begin{align}
\begin{split} \label{showwork}
& \sum_{m = 1}^n \sum_{\{\mathcal{O}_m\}} \int^t \, d t_1 \cdots d t_m\, H ( t_1 ) \cdots H(t_m )  \, c_{\mathcal{O}_m } ( t , t_1 , \dots , t_m )   \left[ \mathcal{O}_{m} \big|_{\xfl}  ( \xvec , t ; t_1 , \dots , t_m )  \right]^{(n)}  = \\
 & \quad \quad \sum_{m = 1}^n \sum_{\{\mathcal{O}_m\}} \sum_{\{\alpha_1, \dots , \alpha_m\}_{\leq n} } c_{\mathcal{O}_m; \alpha_1 , \dots , \alpha_m } (t )  \bar{ \mathbb{C}}^{(n)}_{\mathcal{O}_m; \alpha_1 , \dots , \alpha_m} ( \xvec , t)  \ ,
\end{split}
\end{align}
and then use \eqn{gentimefns} to write the $c_{\mathcal{O}_m; \alpha_1, \dots , \alpha_m}$ as integrals over the intermediate times, so the right-hand side of \eqn{showwork} becomes 
\be
 \sum_{m = 1}^n \sum_{\{\mathcal{O}_m\}} \sum_{\{\alpha_1, \dots , \alpha_m\}_{\leq n} } \int^t \, d t_1 \cdots d t_m H ( t_1 ) \cdots H(t_m )  \, c_{\mathcal{O}_m } ( t , t_1 , \dots , t_m )    \prod_{a = 1}^m  \left( \frac{D(t_a)}{D(t)} \right)^{\alpha_a}  \bar{ \mathbb{C}}^{(n)}_{\mathcal{O}_m; \alpha_1 , \dots , \alpha_m} ( \xvec , t)  \ . 
\ee
Bringing the sum over $\{\alpha_1, \dots , \alpha_m\}_{\leq n} $ inside of the time integrals then leads to \eqn{fulloxflexp}. }
\be \label{fulloxflexp}
 \left[ \mathcal{O}_{m} \big|_{\xfl}  ( \xvec , t ; t_1 , \dots , t_m )  \right]^{(n)}  = \sum_{\{\alpha_1, \dots , \alpha_m\}_{\leq n} } \prod_{a = 1}^m  \left( \frac{D(t_a)}{D(t)} \right)^{\alpha_a}    \bar{\mathbb{C}}^{(n)}_{\mathcal{O}_m; \alpha_1 , \dots , \alpha_m} ( \xvec , t)  \ . 
\ee
The recursion relation is found by taking $d / dt$ of both sides of \eqn{fulloxflexp}.  The right-hand side is easy, and we get 
\begin{align}
\begin{split} \label{rhst}
& \frac{d}{dt }  \sum_{\{\alpha_1, \dots , \alpha_m\}_{\leq n} } \prod_{a = 1}^m  \left( \frac{D(t_a)}{D(t)} \right)^{\alpha_a}    \bar{\mathbb{C}}^{(n)}_{\mathcal{O}_m; \alpha_1 , \dots , \alpha_m} ( \xvec , t)  \\
& \quad \quad = \frac{\dot D(t)}{D(t)} \sum_{\{\alpha_1, \dots , \alpha_m\}_{\leq n} } \left( n - \sum_{i =1 }^m \alpha_i  \right)   \prod_{a = 1}^m  \left( \frac{D(t_a)}{D(t)} \right)^{\alpha_a}   \bar{ \mathbb{C}}^{(n)}_{\mathcal{O}_m; \alpha_1 , \dots , \alpha_m} ( \xvec , t)  \ ,
\end{split}
\end{align}
where we have used that 
\be
\bar{ \mathbb{C}}^{(n)}_{\mathcal{O}_m; \alpha_1 , \dots , \alpha_m} ( \xvec , t)  \propto D(t)^n  \ , 
\ee
which can be seen from \eqn{finalcndef}, using also the fact that $[\mathbb{C}^{ij}_{\mathcal{O}_m, \alpha } ]^{(n)}( \xvec , t ) \propto D(t)^n$ \cite{Donath:2023sav}.  On the left-hand side, we obviously have 
\be
\frac{d}{dt}  \left[ \mathcal{O}_{m} \big|_{\xfl}  ( \xvec , t ; t_1 , \dots , t_m )  \right]^{(n)}  =  \left[ \frac{d}{dt}  \mathcal{O}_{m} \big|_{\xfl}  ( \xvec , t ; t_1 , \dots , t_m )  \right]^{(n)} \ . 
\ee
The expression inside of the brackets is also surprisingly simple.  To see this, consider the product expansion \eqn{oxflprodexp}.  Taking $d / dt$ of the product, we get a sum over terms where $d/dt$ is applied to each factor (i.e. the Leibniz rule).    We can then use \eqn{timederivxflf2} to express the time derivative of each factor as the velocity multiplied by the spatial derivative of the same factor.  Crucially, every term gets the same factor of $v^i$ evaluated at the same point $(\xvec , t)$.   The sum over terms with the spatial derivatives applied to each factor can then be pulled out as the spatial derivative of the product (i.e. reverse Leibniz rule) to give
\begin{align}
\begin{split}
\frac{d}{dt}  \mathcal{O}_{m} \big|_{\xfl}  ( \xvec , t ; t_1 , \dots , t_m ) &  = - \frac{v^i ( \xvec , t )}{a(t)}  \frac{\partial}{\partial x^i }  \mathcal{O}_{m} \big|_{\xfl}  ( \xvec , t ; t_1 , \dots , t_m )  \\
& = \frac{\dot D ( t) }{D(t)}  \frac{\partial_i \theta ( \xvec , t )}{\partial^2}  \frac{\partial}{\partial x^i }  \mathcal{O}_{m} \big|_{\xfl}  ( \xvec , t ; t_1 , \dots , t_m )  \ ,
\end{split}
\end{align}
where in the last line we assumed zero vorticity.  This gives
\begin{align}
\begin{split}
\frac{d}{dt}  \left[ \mathcal{O}_{m} \big|_{\xfl}  ( \xvec , t ; t_1 , \dots , t_m )  \right]^{(n)}  & =  \frac{\dot D ( t) }{D(t)} \left[ \frac{\partial_i \theta ( \xvec , t )}{\partial^2}  \frac{\partial}{\partial x^i }  \mathcal{O}_{m} \big|_{\xfl}  ( \xvec , t ; t_1 , \dots , t_m ) \right]^{(n)} \\
& =  \frac{\dot D ( t) }{D(t)}  \sum_{\ell = 1 }^{n-1} \frac{\partial_i \theta^{(n - \ell) }( \xvec , t )}{\partial^2}  \frac{\partial}{\partial x^i }  \left[ \mathcal{O}_{m} \big|_{\xfl}  ( \xvec , t ; t_1 , \dots , t_m )  \right]^{(\ell)} \\
& \hspace{-1.5in} =  \frac{\dot D ( t) }{D(t)}  \sum_{\ell = 1 }^{n-1} \sum_{\{\alpha_1, \dots , \alpha_m\}_{\leq \ell} }   \left[  \prod_{a = 1}^m  \left( \frac{D(t_a)}{D(t)} \right)^{\alpha_a}    \right]\frac{\partial_i \theta^{(n - \ell) }( \xvec , t )}{\partial^2}  \partial_i  \bar{\mathbb{C}}^{(\ell)}_{\mathcal{O}_m; \alpha_1 , \dots , \alpha_m} ( \xvec , t)  \ , 
\end{split}
\end{align}
where we used \eqn{fulloxflexp} to plug in for $ \left[ \mathcal{O}_{m} \big|_{\xfl}  ( \xvec , t ; t_1 , \dots , t_m )  \right]^{(\ell)} $.  Now, to match with \eqn{rhst}, we want to swap the order of the sums to get
\begin{align}
\begin{split}
& \frac{d}{dt}  \left[ \mathcal{O}_{m} \big|_{\xfl}  ( \xvec , t ; t_1 , \dots , t_m )  \right]^{(n)}  \\
& \quad \quad  =  \frac{\dot D ( t) }{D(t)}   \sum_{\{\alpha_1, \dots , \alpha_m\}_{\leq n-1} } \left[ \prod_{a = 1}^m  \left( \frac{D(t_a)}{D(t)} \right)^{\alpha_a}  \right] \sum_{\ell = \alpha_1 + \dots + \alpha_m }^{n-1}  \frac{\partial_i \theta^{(n - \ell) }( \xvec , t )}{\partial^2}  \partial_i   \bar{\mathbb{C}}^{(\ell)}_{\mathcal{O}_m; \alpha_1 , \dots , \alpha_m} ( \xvec , t)  \ . 
\end{split}
\end{align}
Since this has to be equal to \eqn{rhst} for arbitrary times $t_a$, we can match powers of $D(t_a)$, which gives 
\be \label{newxflrec}
 \left( n - \sum_{i =1 }^m \alpha_i  \right)      \bar{\mathbb{C}}^{(n)}_{\mathcal{O}_m; \alpha_1 , \dots , \alpha_m} ( \xvec , t) =  \sum_{\ell = \alpha_1 + \dots + \alpha_m }^{n-1}  \frac{\partial_i \theta^{(n - \ell) }( \xvec , t )}{\partial^2}  \partial_i   \bar{\mathbb{C}}^{(\ell)}_{\mathcal{O}_m; \alpha_1 , \dots , \alpha_m} ( \xvec , t) \ ,
\ee
for $\sum_{i =1 }^m \alpha_i  < n$.  So, all of the kernels with $\sum_{i =1 }^m \alpha_i  < n$ are in this way determined by lower-order kernels.  

This leave us to find the kernels with $\sum_{i =1 }^m \alpha_i  =  n$.   Looking back at the definition of the kernels \eqn{finalcndef}, we see that these are actually the simplest ones to form.  In \eqn{finalcndef}, we sum over all the tuples $(n_1 , \dots , n_m )$ such that $n_1 + \cdots + n_m = n$ and $n_a \geq \alpha_a$ for $a = 1 , \dots , m$.  For the terms with $\sum_{i =1 }^m \alpha_i  =  n$, the only possibility is to have $n_a = \alpha_a$ for $a = 1 , \dots , m$, so there is only one term in the sum
\be \label{finrec}
\bar{\mathbb{C}}^{(n)}_{\mathcal{O}_m; \alpha_1 , \dots , \alpha_m}  ( \xvec , t) \Big|_{\sum_{i =1 }^m \alpha_i  =  n} =      \Delta_{\mathcal{O}_m} [ i_1 , j_1 , \dots , i_m , j_m ]   \prod_{a = 1}^m   [ \mathbb{C}_{ (\mathcal{O}_m , a)  , \alpha_a}^{ i_a j_a}]^{(\alpha_a)} ( \xvec , t )    \ .
\ee
As an example of this relation, consider the operator $\mathcal{O}_2 = r^{ij} p^{ij}$.  For $n = 4$, there are three terms with $\alpha_1 + \alpha_2 = 4$, i.e. $(\alpha_1 , \alpha_2) = (3,1) , (2,2) , (1,3)$.  Using \eqn{finrec}, these are simply given by
\begin{align}
\begin{split}
\bar{\mathbb{C}}^{(4)}_{[rp]; 3 , 1 } =  [\mathbb{C}^{ij}_{r,3}]^{(3)} [\mathbb{C}^{ji}_{p,1}]^{(1)}  \ , \quad \bar{\mathbb{C}}^{(4)}_{[rp]; 2 , 2 } =  [\mathbb{C}^{ij}_{r,2}]^{(2)} [\mathbb{C}^{ji}_{p,2}]^{(2)}  \andd \bar{\mathbb{C}}^{(4)}_{[rp]; 1 , 3 } =  [\mathbb{C}^{ij}_{r,1}]^{(1)} [\mathbb{C}^{ji}_{p,3}]^{(3)}   \ . 
\end{split}
\end{align}

Thus, we have been lead to a much more efficient way to compute all of the perturbative kernels.  For an operator $\mathcal{O}_m$ at order $n$, all of the kernels $\bar{\mathbb{C}}^{(n)}_{\mathcal{O}_m; \alpha_1 , \dots , \alpha_m} $ with $\sum_{i =1 }^m \alpha_i  <  n$ can be determined from lower order kernels using \eqn{newxflrec}, and all of the kernels with $\sum_{i =1 }^m \alpha_i  =  n$ can be formed using the simple formula \eqn{finrec}.   An additional relationship satisfied by our kernels is the equal-time completeness relation
 \be \label{equaltimerltn}
\mathcal{O}_m^{(n)} ( \xvec , t)  = \sum_{\{\alpha_1, \dots , \alpha_m\}_{\leq n} }     \bar{\mathbb{C}}^{(n)}_{\mathcal{O}_m; \alpha_1 , \dots , \alpha_m} ( \xvec , t)   \ ,
\ee
which can be easily obtained by setting $t_1 = \cdots = t_m = t$ in \eqn{fulloxflexp}, where $\mathcal{O}_m^{(n)}$ is the local-in-time contraction, \eqn{omfactors}, of $r^{ij}$ and $p^{ij}$ at $n$-th order in terms of the standard $F_i$ and $G_i$ kernels.

%
%
\section{Sixth-order galaxy bias} \label{sixordersec}

In \cite{Donath:2023sav}, it was shown that at fifth order in the galaxy bias, the non-local-in-time expansion has three more bias parameters than the local-in-time expansion ($29$ and $26$ respectively), meaning that the galaxy clustering signal at a fixed redshift is sensitive to the formation time of galaxies.\footnote{ Non-local-in-time effects have also been illuminated in other areas, such as halo assembly bias \cite{Gao:2005ca, Croton:2006ys, Mao:2017aym}, galaxy intrinsic alignments \cite{Schmitz:2018rfw}, and the Lyman-$\alpha$ forest and general selection effects \cite{Ansari:2024efj}. } In \cite{Donath:2023sav} (as well as in earlier related works), to arrive at this result, it was enough to consider a subset of non-local-in-time composite operators, specifically ones where all fields $r^{ij}$ and $p^{ij}$ are evaluated at the same past time of the fluid trajectory.   However, as mentioned, this is not the most general expansion of $N_g$ in \eqn{nggeneq}, and starting at sixth order, it becomes necessary to use the most general expansion.

The procedure outlined above, while ensuring that we consider all possible functional forms, generally produces an overcomplete set of functions at a given order.  Thus, at a fixed order, we can look for degeneracies among the set of functions and find the minimal basis.    This is the procedure followed in \cite{Angulo:2015eqa, DAmico:2022osl, DAmico:2022ukl, Donath:2023sav, Ansari:2024efj, Anastasiou:2025jsy}, for example.  At sixth order, this starts to become too great a task for a naive implementation in Mathematica.  To address this, we have built a custom package in Rust that employs a number of
optimizations, including the use of finite field methods \cite{Peraro:2016wsq}
to perform exact numerical analysis using machine-word-size integers, the
tracking of stabilizer/coset permutations to reduce the number of explicit terms
that are tracked and operated on, and aggressive parallelization via the Rust
Rayon library.\footnote{The package is capable of constructing the six point basis
discussed below in $\sim 5$ minutes using 16 threads on a laptop with a Ryzen 7 7840U
CPU, and we expect there are further optimizations available.  We plan to
release the package as part of a future publication, but a preliminary version
is available upon request.}  Other ways of constructing bias bases have been given in \cite{Mirbabayi:2014zca, Desjacques:2016bnm, Eggemeier:2018qae, Fujita:2020xtd, DAmico:2021rdb, Eggemeier:2021cam, Schmidt:2020ovm, Marinucci:2024add, Ansari:2025nsf}.  However, we believe that our procedure makes the non-local-in-time structure of galaxy clustering more manifest and ensures that we consider all possible shapes in the galaxy clustering signal.

As mentioned previously and discussed in \appref{vorticityapp}, we can ignore vorticity for our purposes.  This means that $p^{ij}$ is symmetric and given by \eqn{pijpert}.  If vorticity is included, then $p^{ij}$ is no longer symmetric, and one should include both $p^{ij}$ and its transpose $p^{ji} \equiv (p^{ij})^t$ (or equivalently the symmetric and antisymmetric parts of $p^{ij}$) when constructing all possible scalar contractions.  However, we now proceed assuming $p^{ij}$ is symmetric.

%
%
\subsection{Basis of descendants}

Here, we give our results in the basis of descendants \cite{Angulo:2015eqa}, where if $\mathbb{C}_{\mathcal{O}_m; \alpha_1 , \dots , \alpha_m}^{(n)}$ is used in the basis at order $n$, then $\mathbb{C}_{\mathcal{O}_m; \alpha_1 , \dots , \alpha_m}^{(n+1)}$ is used in the basis at order $n+1$.\footnote{There is a typo at the beginning of App. C in \cite{Donath:2023sav} for the definition of the basis of descendants, although implementation and other equations, including Eq. C2, are correct. The beginning of App. C should say that if $\mathbb{C}^{(n)}_{\mathcal{O}_m, \alpha}$ is used at order $n$, then $\mathbb{C}^{(n+1)}_{\mathcal{O}_m, \alpha}$ is used at order $n+1$.}  Calling  the independent elements of the basis of descendants $\mathbb{B}^{(n)}_i$, this means that the basis at each order can be expressed as the union
\be
\{ \mathbb{B}^{(n)} \} = \{ \mathbb{B}^{(n-1)} \}\Big|_{\bar{\mathbb{C}}^{(n-1)}_{\mathcal{O}_m ; \alpha_1 , \dots , \alpha_m}  \rightarrow \bar{\mathbb{C}}^{(n)}_{\mathcal{O}_m ; \alpha_1 , \dots , \alpha_m} }  \cup  \Delta \{ \mathbb{B}^{(n)} \}  \ ,
\ee
where $\{ \mathbb{B}^{(n-1)} \}\Big|_{\bar{\mathbb{C}}^{(n-1)}_{\mathcal{O}_m ; \alpha_1 , \dots , \alpha_m}  \rightarrow \bar{\mathbb{C}}^{(n)}_{\mathcal{O}_m ; \alpha_1 , \dots , \alpha_m} }  $ is the basis of descendants at order $n-1$, but with all of the $\bar{\mathbb{C}}^{(n-1)}_{\mathcal{O}_m ; \alpha_1 , \dots , \alpha_m}$ replaced by  $\bar{\mathbb{C}}^{(n)}_{\mathcal{O}_m ; \alpha_1 , \dots , \alpha_m}$, and $\Delta \{ \mathbb{B}^{(n)} \} $ are the new basis elements at order $n$.  We find a basis of descendants up to sixth order to be
\begin{align}
\begin{split} \label{bod1}
& \{ \mathbb{B}^{(1)} \} = \{\mathbb{C}_{\delta,1}^{(1)}   \} \ , \quad  \Delta \{ \mathbb{B}^{(2)} \} = \{  \mathbb{C}_{\delta,2}^{(2)}, \mathbb{C}_{\delta^2,1}^{(2)}   \}  \ , \quad \Delta \{ \mathbb{B}^{(3)} \} = \{  \mathbb{C}_{\delta,3}^{(3)}, \mathbb{C}_{\delta^2,2}^{(3)}, \mathbb{C}_{[r^2],2}^{(3)}, \mathbb{C}_{\delta^3,1}^{(3)}   \}   \ , \\
& \Delta \{ \mathbb{B}^{(4)}  \} = \{ \mathbb{C}_{\delta,4}^{(4)}, \mathbb{C}_{\delta^2,3}^{(4)}, \mathbb{C}_{[r^2],3}^{(4)}, \mathbb{C}_{[r^3],2}^{(4)}, \mathbb{C}_{\delta^3,2}^{(4)}, \mathbb{C}_{[r^2]\delta,2}^{(4)}, \mathbb{C}_{\delta^4,1}^{(4)}, \mathbb{C}_{ [ r^3] \delta,1}^{(4)}  \}  \ ,  \\
& \Delta \{ \mathbb{B}^{(5)}  \} = \{ \mathbb{C}_{ \delta , 5 }^{(5)} , \mathbb{C}_{ \delta^2 , 4 }^{(5)}, \mathbb{C}_{ [r^2] ,4 }^{(5)} , \mathbb{C}_{ \delta^3 , 3 }^{(5)}, \mathbb{C}_{ [r^3] , 3 }^{(5)} , \mathbb{C}_{ [r^2] \delta , 3 }^{(5)} , \mathbb{C}_{ \delta^4 , 2 }^{(5)}  , \\
& \hspace{1in}    \mathbb{C}_{ [r^3] \delta,2 }^{(5)} , \mathbb{C}_{ [r^4] ,2 }^{(5)}, \mathbb{C}_{ \delta^5,1 }^{(5)} , \mathbb{C}_{ [r^5] ,1  }^{(5)}, \mathbb{C}_{ [r^4] \delta,1 }^{(5)} , \mathbb{C}_{ [r^3] \delta^2,1 }^{(5)} , \mathbb{C}_{ [p^3],3 }^{(5)}    \}  \ ,  \\
& \Delta \{ \mathbb{B}^{(6)}  \} = \{
 \mathbb{C}^{(6)}_{\delta,6},
\mathbb{C}^{(6)}_{\delta^{2},5},
\mathbb{C}^{(6)}_{\delta^{3},4},
\mathbb{C}^{(6)}_{\delta^{4},3},
\mathbb{C}^{(6)}_{\theta^{2},5},
\mathbb{C}^{(6)}_{[p^{2}]^2,3}, 
 \mathbb{C}^{(6)}_{[r^{2}],5}, 
\mathbb{C}^{(6)}_{[r^{2}]\delta,4},
\mathbb{C}^{(6)}_{[r^{2}]\delta^{2},3},
\mathbb{C}^{(6)}_{[r^{2}]\delta^{3},2}, 
\mathbb{C}^{(6)}_{[r^{2}]\theta,4}, \\
& \hspace{1in} \mathbb{C}^{(6)}_{[r^{2}]^2,3},  \mathbb{C}^{(6)}_{[r^{2}]^2 \delta,2},
\mathbb{C}^{(6)}_{[r^{2}]^2 \delta^{2},1},
\mathbb{C}^{(6)}_{[p^{3}],4},
\mathbb{C}^{(6)}_{[p^{3}]\theta,3},
\mathbb{C}^{(6)}_{[p r^{2}],4}, 
 \mathbb{C}^{(6)}_{[r^{3}],4},
\mathbb{C}^{(6)}_{[r^{3}]\delta,3},
\mathbb{C}^{(6)}_{[r^{3}]\delta^{2},2},\\
& \hspace{1in}  \mathbb{C}^{(6)}_{[r^{2}][r^{3}],2},
\mathbb{C}^{(6)}_{[r^{2}][r^{3}]\delta,1},
\mathbb{C}^{(6)}_{[r^{3}]^{2},1},
\mathbb{C}^{(6)}_{[p^{2} r^{2}],3},
\mathbb{C}^{(6)}_{[r^{4}]\delta,2},
\mathbb{C}^{(6)}_{[r^{2}][r^{4}],1},
\mathbb{C}^{(6)}_{[r^{5}]\delta,1},
\bar{\mathbb{C}}^{(6)}_{[r^{2}];1,5} \} \ .
\end{split}
\end{align}
Thus, we find a total of $57$ bias parameters at sixth order, $56$ of which can, for simplicity, be written in the expansion of \cite{Donath:2023sav} (hence the appearance of mostly the unbarred $\mathbb{C}^{(n)}$, which are given in terms of the kernels derived in this work in \eqn{heretobareq}).  However, the basis of \cite{Donath:2023sav} is not complete at sixth order, and we must add one kernel derived in this work, which we take to be $\bar{\mathbb{C}}^{(6)}_{[r^{2}];1,5}$.  We have attached explicit expressions for all of the kernels in \eqn{bod1} in ancillary files associated with the arXiv preprint of this paper.

This then allows us to express the $n$-th order galaxy overdensity as
\be \label{bodgenexpand}
\delta_g^{(n)} ( \xvec , t ) = \sum_{i = 1 }^{N_n} c_i ( t ) \mathbb{B}^{(n)}_i ( \xvec , t )  \ , 
\ee
where $N_n$ is the total number of independent basis functions at order $n$ (i.e. $N_1 = 1$, $N_2 = 3$, $N_3 = 7$, $N_4 = 15$, $N_5 = 29$, $N_6 = 57$), $c_i$ are the bias parameters, and the ordering of the $\mathbb{B}^{(n)}_i $ is inherited by the ordering in \eqn{bod1}.  In terms of the $b_i$ parameters used in previous works (see e.g. \cite{DAmico:2022ukl, Donath:2023sav}), we have
\be
\{ c_1 , \dots , c_{10} \} = \{ b_1 , b_2 , b_5, b_3, b_6, b_8, b_{10}, b_4 , b_7 , b_9 \}  \ ,
\ee
and $c_i = b_i$ for $i = 11, \dots , 57$.  Notice in \eqn{bodgenexpand} that once a bias parameter $c_i$ appears at some order, it appears at all higher orders; the fluid expansion ensures that these appear in a consistent way so that the final result is an NRNGP scalar, and the relationship that it induces among kernels of different orders is the basis of the soft limits that we derive in \secref{plimsec} and \secref{kerlimitssec}.  

Similar to what was done in \cite{Donath:2023sav}, we can compare our sixth order non-local-in-time basis to a local-in-time basis at sixth order.  The local-in-time basis is found by using \eqn{generaltauij} and setting the time dependent response functions $c_{\mathcal{O}_m}(t, t_1 , \dots , t_m)$ proportional to Dirac delta functions which set all times equal to the final time $t$, see \eqn{litcoeffdeltafn}.  Upon integration over the times $t_1, \dots , t_m$, this forces $\xfl( \xvec , t , t_i) \rightarrow \xvec$ so that one ends up with a local expansion of the galaxy overdensity $\delta_{g, \text{loc.}}$
\be
\delta_{g, \text{loc.}}^{(n)} ( \xvec , t ) = \sum_{m=1}^n \sum_{\{ \mathcal{O}_m \}} c_{\mathcal{O}_m} ( t ) \mathcal{O}_m^{(n)} ( \xvec , t )  \ , 
\ee
where $ c_{\mathcal{O}_m} ( t ) $ are given in \eqn{litcoeffdeltafn}, and $\mathcal{O}_m^{(n)}$ are just the local-in-time contractions of $r^{ij} ( \xvec , t)$ and $p^{ij} ( \xvec , t)$.   One can then find a minimal basis at each order $n$.  In \cite{Donath:2023sav}, it was found that at fifth order, the local-in-time basis has 26 elements, as opposed to the non-local-in-time basis which has 29 elements.  This implies that the following combinations of bias parameters are zero for local-in-time structure formation
\begin{align}
\begin{split} \label{litrel1}
& 0 = b_1 - 4 b_2 + 6 b_3 - 4 b_4 + 90 b_8 - 76 b_9 + b_{16}  \ , \\
\text{local-in-time} \Rightarrow \hspace{0.5in}  & 0 = b_{18} - b_{9}   \ , \\ 
& 0 =  - \frac{4 b_8}{3} + \frac{4 b_9 }{3} - \frac{10 b_{11}}{3} + \frac{7 b_{20}}{3} + b_{29}  \ .
\end{split}
\end{align}

We now do the same analysis at sixth order.  We have found that a local-in-time basis at sixth order contains 46 elements, which we can take to be
\begin{align}
\begin{split} \label{augmented1}
& \{
\theta,\;
\theta [p^{2}],\;
\theta [p^{2}][p r],\;
\delta\theta [p^{2}],\;
\delta^{2}\theta [p^{2}],\;
\theta [p^{2}][ r^{2}],\;
\theta [p^{3}],\;
\delta\theta [p^{3}],\;
\theta [p^{4}],\;
\theta [p^{3} r],\; \\
& \quad  \theta [p^{2} r],\; 
 \delta\theta [p^{2} r],\; 
\theta [p r],\;
\delta\theta [p r],\;
\delta^{2}\theta [p r],\; 
 \theta [p r^{2}],\;
\delta\theta,\;
\delta\theta [r^{2}],\;
\delta^{2}\theta,\;
\delta^{3}\theta,\;
\theta [r^{2}],\; \\
&  \quad \theta^{2},\;
\theta^{2}[p^{2}],\;
\delta\theta^{2}[p^{2}],\; 
 \theta^{2}[p^{3}],\; 
\delta\theta^{2},\;
 [p^{2}],\;
[p^{2}][p^3],\;
[p^{2}][p^2 r],\;
[p^{2}][p r],\;
\delta [p^{2}][p r],\;
\delta [p^{2}],\;
[p^{2} ][r^{2}],\; \\
&  \quad [p^{2}]^{2},\; 
[p^{3}],\;
[p r ][p^{3}],\;
\delta[p^{3}],\;
[p^{2} r],\; 
 [p^{2} r^{2}],\;
[p r],\; 
 \delta[p r],\;
[p r^{2}],\;
\delta,\;
\delta [r^{2}][r^3],\;
[r^{2}],\;
[r^{3}]
\} \ . 
\end{split}
\end{align}
This local-in-time basis can be augmented by the following elements to make a basis that is equivalent to the full 57 element non-local-in-time basis of descendants given in \eqn{bod1}
\begin{align}
\begin{split} \label{augmented2}
\{ \mathbb{C}_{\delta , 1}^{(6)},  \mathbb{C}_{\delta , 2}^{(6)},  \mathbb{C}_{\delta^2 , 1}^{(6)},  \mathbb{C}_{[r^2] , 2}^{(6)} ,  \mathbb{C}_{[r^2] , 3}^{(6)},  \mathbb{C}_{[r^2] , 4}^{(6)} ,  \mathbb{C}_{[r^2] \delta  , 2}^{(6)} ,  \mathbb{C}_{[r^2]\delta , 3}^{(6)},  \mathbb{C}_{[r^3] , 2}^{(6)},  \mathbb{C}_{[r^3] , 3}^{(6)} , \bar{\mathbb{C}}_{[r^2] ; 1,5}^{(6)} \} \ . 
\end{split}
\end{align}
Thus, if we change basis from the basis of descendants \eqn{bod1} to the augmented basis \eqn{augmented1} and \eqn{augmented2}, the coefficients of the operators \eqn{augmented2} have to be zero for local-in-time structure formation.  Combining with the conditions \eqn{litrel1}, this means that the following combinations of bias parameters are zero if structure formation were local in time
\begin{align}
0 &= -b_{46} - 3 b_{47} + 3 b_{20}, \  \nonumber \\[6pt]
0 &= -b_{45} - \frac{8}{3} b_{47} - \frac{7}{3} b_{48}
     - \frac{20}{9} b_{53} + \frac{4}{3} b_{13} - b_{15} \nonumber \\
& \quad  \quad    + \frac{28}{15} b_{20} - \frac{4}{3} b_{21} + \frac{10}{3} b_{23}
     + \frac{40}{9} b_{24} + \frac{4}{5} b_{29}
     - \frac{16}{15} b_{8} + \frac{16}{15} b_{9} \ , \nonumber \\[6pt] 
\text{local-in-time} \Rightarrow \quad \quad 0 &= -b_{44} + 4 b_{47} - \frac{7}{2} b_{20}
     + \frac{3}{2} b_{29} + \frac{2}{3} b_{8} - \frac{2}{3} b_{9} \ , \\[6pt]
0 &= -b_{37} - b_{40} + b_{21} \ , \nonumber \\[6pt] 
0 &= -b_{36} + b_{9} \ ,  \nonumber \\[6pt] 
0 &= -b_{31} - 5 b_{34} - 45 b_{13} + 4 b_{17} + 38 b_{21}
     - b_{5} + 4 b_{6} - 6 b_{7} - 35 b_{8} + \frac{65}{2} b_{9} \ , \nonumber \\[6pt] 
0 &= -b_{30} + 4 b_{16} - b_{2} + 4 b_{3} - 6 b_{4}
     + 60 b_{8} - \frac{113}{2} b_{9} \ ,  \nonumber \\
     0 & = b_{57} \nonumber \ . 
\end{align}
Said another way, measurement of an order one value for any of the above combinations of bias parameters is a direct measurement that galaxies and dark-matter halos form on time scales of order Hubble.  We have attached explicit expressions for all of the kernels in \eqn{augmented1} and \eqn{augmented2} in ancillary files associated with the arXiv preprint of this paper.

%
%
\subsection{$p$-particle soft limits} \label{plimsec}

In this section, we show that our kernels, and ultimately the non-local-in-time galaxy overdensity $\delta_g$, satisfy a series of multi-leg soft limits (which we call $p$-particle soft limits) that arise when the sum of a subset of the external momenta goes to zero.  These limits were pointed out in \cite{DAmico:2021rdb}, and here we explicitly show that they are satisfied by our non-local-in-time construction (in the EdS approximation with no velocity vorticity).  We provide an explicit and different proof from \cite{DAmico:2021rdb}, which allows us to give a precise procedure for taking the $p$-particle soft limits.

Here we give a brief sketch of our proof, which is presented in detail in \appref{plimapp}.  We find it easiest to first work at the level of the fields in position space, as opposed to the kernels in Fourier space, since we have all of the expressions explicitly in terms of $\delta$ and $\theta$, as will become evident below.  We start by noticing that, ultimately, the expression for the perturbations of any scalar $X$, including the galaxy overdensity $\delta_g$ and $\delta$ and $\theta$ themselves, can be written as a function of the fields as
\be \label{xngendef}
X^{(n)} \left[ \partial_i , \delta^{ij}_K  ,  \left\{ \frac{\partial_i \partial_j \delta^{(a)}}{\partial^2} , \frac{\partial_i \partial_j \theta^{(a)}}{\partial^2}  , \frac{\partial_i \theta^{(a)}}{\partial^2}  \right\}_{a = 1 , \dots , n} \right] \ ,
\ee
as it follows from the iterative solution for the kernels.
For the fields $\delta$ and $\theta$, this form is obvious from the perturbative solutions to the equations of motion \eqn{dmeomanbn} (although only perturbations up to $a = n-1$ appear), while for general non-local-in-time scalars, it is a consequence of the fluid expansion.  In terms of lower order fields, each of the building blocks of \eqn{xngendef} are not necessarily made simply of products of the fields shown in \eqn{xngendef}, but rather there can be non-local operators acting on products of the perturbations, as in, for example
\be
\frac{\partial_i \partial_j \delta^{(4)} }{\partial^2} \supset \frac{\partial_i \partial_j  }{\partial^2} \left( \theta^{(1)} \partial_k \delta^{(1)} \frac{\partial_k \theta^{(2)}}{\partial^2} \right)  \ . 
\ee
The $p$-particle soft limit is roughly the limit when a sum of $p$ external momenta goes to zero, i.e. $\qvec_1 + \dots + \qvec_p \rightarrow 0$ (although we make this more precise below).  From \eqn{xngendef}, we see that in this limit, there is a special term, $\partial_i \theta^{(p)} / \partial^2$, that would appear to have a pole going like $\sim \mathcal{O}( | \qvec_{1;p}|^{-1} ) $, but because of dark-matter mass and momentum conservation, we have $\theta^{(p)} \sim \mathcal{O}( | \qvec_{1;p}|^{2} )$, so in fact there is no actual pole.  This term is still special, though, because it is the only one that is allowed to appear with a $\partial_i / \partial^2$, i.e. with only one spatial derivative above the Laplacian. So, since we have $X^{(n)}$ directly in terms of the fields, we can simply explicitly track all of the terms with $\partial_i \theta^{(p)} / \partial^2$ through the perturbative solution.   In field theoretic language, this corresponds to taking the limit $\qvec_{1;p} \rightarrow 0$ while at the same time keeping $\theta^{(p)}$ \emph{off shell}, that is, being ignorant about the explicit solution for $\theta^{(p)}$ from the equations of motion.

We can now give our definition of the $p$-particle soft limit (which we denote as $\underset{p}{\rightarrow}$) in this setup.   First, we explicitly collect all of the terms with $\partial_i \theta^{(p)} / \partial^2$ in \eqn{xngendef}.  Then we send the total momentum of $\theta^{(p)}$ to zero, which at the level of the fields means that we can pull $\partial_i \theta^{(p)} / \partial^2$ out of other derivatives, for example
\be
\partial_j \left( \partial_i \delta^{(a)} \frac{\partial_i \theta^{(p)} }{\partial^2} \right) \approx  \frac{\partial_i \theta^{(p)} }{\partial^2} \partial_j  \partial_i \delta^{(a)}  \ . 
\ee
In \appref{dmplimapp}, using this definition and the dark-matter equations of motion, we first show that the dark-matter fields satisfy
\begin{align}
\begin{split} \label{dmirlimits2}
& \delta^{(n)} \underset{p}{\rightarrow} \frac{1}{p} \partial_i \delta^{(n-p)} \frac{\partial_i \theta^{(p)} }{\partial^2}  \andd  \theta^{(n)} \underset{p}{\rightarrow} \frac{1}{p} \partial_i \theta^{(n-p)} \frac{\partial_i \theta^{(p)} }{\partial^2} \ ,
\end{split}
\end{align}
for $1 \leq p \leq n-1$.  In \appref{dmplimfluidapp}, we use this to prove some intermediate results about the fluid expanded kernels $\mathbb{C}_{\delta , \alpha}^{(n)}$ and $\mathbb{C}_{\theta , \alpha}^{(n)}$.  In \appref{multiplim} we show that the non-local-in-time kernels that have $\sum_{i =1 }^m \alpha_i  <  n$ satisfy
\be
\bar{ \mathbb{C}}^{(n)}_{\mathcal{O}_m; \alpha_1 , \dots , \alpha_m}  \underset{p }{\rightarrow}  \frac{1}{p}  \partial_i  \bar{ \mathbb{C}}^{(n-p)}_{\mathcal{O}_m; \alpha_1 , \dots , \alpha_m} \frac{\partial_i \theta^{(p)}}{\partial^2}  \ ,
\ee
for $1 \leq p \leq n - \sum_{i =1 }^m \alpha_i $, and
\be
\bar{\mathbb{C}}^{(n)}_{\mathcal{O}_m; \alpha_1 , \dots , \alpha_m} \Big|_{\sum_{i =1 }^m \alpha_i  =  n} \underset{p }{\rightarrow}  0 \ .
\ee
This allows us to finally conclude that
\be \label{deltagplimmain}
\delta^{(n)}_g \underset{p }{\rightarrow}  \frac{1}{p} \partial_i \delta^{(n-p)}_g \frac{\partial_i \theta^{(p)}}{\partial^2} \ ,
\ee
for $1 \leq p \leq n-1$.

%
%
\subsection{$p$-particle soft limits of Fourier-space kernels} \label{kerlimitssec}

From \secref{plimsec}, we have an explicit procedure at the level of the fields to prove the $p$-particle soft limits.  This now allows us to give an unambiguous procedure for doing the same at the level of the Fourier-space kernels.  The final perturbative expression \eqn{bodgenexpand} for the galaxy overdensity can be written in terms of the kernels $K_n$, which also depend on the bias parameters according to \eqn{bodgenexpand}, in the following way
\begin{align}
\begin{split} \label{biaskernels}
\delta^{(1)}_g ( \kvec , a ) & = D(a) K_1 ( \kvec) \tilde \delta_{\kvec}^{(1)}  \ ,\\
\delta^{(n)}_g ( \kvec , a ) & = D(a)^n \int^{\kvec}_{\kvec_1 , \dots , \kvec_n} K_n( \kvec_1 , \dots , \kvec_n) \tilde \delta^{(1)}_{\kvec_1} \cdots \tilde \delta^{(1)}_{\kvec_n} \ ,
\end{split}
\end{align}
for $n \geq 2$.   It is well known (as a consequence of the LSS consistency conditions  \cite{Kehagias:2013yd, Creminelli:2013mca, Creminelli:2013poa, Horn:2014rta, Creminelli:2012ed, Hinterbichler:2012nm, Hinterbichler:2013dpa}, for example) that  the perturbative kernels of the galaxy overdensity satisfy the soft limit\footnote{These relationships also apply using exact time dependence and for certain dark-energy theories \cite{Lewandowski:2017kes, Crisostomi:2019vhj, Lewandowski:2019txi}.   } 
\begin{align}
\begin{split} \label{k1kerlims}
\lim_{\qvec_1  \rightarrow 0}  K_n ( \qvec_1 , \kvec_2 , \dots , \kvec_n ) =  & \frac{1}{n} \frac{\qvec_1 \cdot  \kvec_{2;n} }{q_1^2} K_{n-1} ( \kvec_{2}, \dots , \kvec_n )  \ ,
\end{split}
\end{align}
which is the $p = 1$ case of \eqn{deltagplimmain}, 
\be \label{k1kerlimsps}
\delta_g^{(n)}  \underset{1}{\rightarrow} \partial_i \delta^{(n-1)}_g \frac{\partial_i \theta^{(1)}}{\partial^2} \ .
\ee
This is the simplest case, since there really is a pole going like $\sim \mathcal{O}(q_1^{-1})$ when $\qvec_1 \rightarrow 0$.  

The $p\geq2$ limits, on the other hand, require a more sophisticated analysis.  From \secref{plimsec}, we obtained the $p$-particle soft limits by first tracking the factor of $\partial_i \theta^{(p)} / \partial^2$, and then taking the total momentum of $\theta^{(p)}$ to zero.  At the level of the kernels, on the other hand, where the velocity is on shell and we have written the solution \eqn{biaskernels} in terms of the lowest order field $\tilde \delta^{(1)}_{\kvec_i}$, we cannot simply track the factors of $\partial_i \theta^{(p)} / \partial^2$.  However, as we will show, we can reproduce the procedure of \secref{plimsec} by first sending just the \emph{magnitude} $|\qvec_{1;p}| \rightarrow 0$, and then sending the total momentum $\qvec_{1;p} \rightarrow 0$.  To do that precisely, we first define exactly how we will take those limits.  

 Consider a  function $f ( \kvec_1 , \dots , \kvec_n)$ which is a sum of rational functions of dot products of the vector arguments $( \kvec_1 , \dots , \kvec_n)$ (this is the form of all of the kernels that we consider here).  We define the vector limits $\mathcal{V}_{\kvec_i \rightarrow 0}^{(\ell)} [ f( \kvec_1 , \dots , \kvec_n)] $ in terms of the formal series expansion
\be
f ( \kvec_1 , \dots, \lambda \, \kvec_i , \dots  , \kvec_n) = \sum_\ell \lambda^\ell \mathcal{V}_{\kvec_i \rightarrow 0}^{(\ell)}[ f( \kvec_1 , \dots , \kvec_n)] \ .
\ee
We can also define a limit where we only send the magnitude of some vector to zero in a specific way.  Since $f( \kvec_1 , \dots , \kvec_n)$ is made of dot products of its vector arguments, we can also write 
\be
f( \kvec_1 , \dots , \kvec_n) = f( \{\kvec_i \cdot \kvec_j\}_{i \neq j} ; |\kvec_1| , \dots , | \kvec_n| ) \ ,
\ee
that is, we expand the dot products of sums of vectors to write the function explicitly in terms of only dot products of a pair of the vector arguments.  Then we can define the magnitude limits $\mathcal{M}_{|\kvec_i| \rightarrow 0}^{(\ell)} [ f( \kvec_1 , \dots , \kvec_n)] $ in terms of the formal series expansion
\be
f( \{\kvec_i \cdot \kvec_j\}_{i \neq j} ; |\kvec_1| , \dots , \lambda |\kvec_i| , \dots , | \kvec_n| ) = \sum_\ell \lambda^\ell \mathcal{M}_{|\kvec_i| \rightarrow 0}^{(\ell)}[ f( \kvec_1 , \dots , \kvec_n)] \ .
\ee

This leads us to the specific limiting procedure that we will use on the $K_n$ kernels.  Letting $\qvec_{1;p}$ be the momentum that we will send to zero in the $p$-particle limit, we first replace $\qvec_p = \qvec_{1;p} - \qvec_{1;p-1}$, 
\be
K_n ( \qvec_1 , \dots , \qvec_{p-1}, \qvec_{1;p} - \qvec_{1;p-1} , \kvec_{p+1} , \dots , \kvec_n )  \ ,
\ee
which is now a function of $(\qvec_1 , \dots , \qvec_{p-1}, \qvec_{1;p},  \kvec_{p+1} , \dots , \kvec_n)$.  Then we take the magnitude limit and take the piece that is proportional to $| \qvec_{1;p}|^{-2}$, i.e.
\be \label{maglim1}
 \mathcal{M}^{(-2)}_{ |\qvec_{1;p}| \rightarrow 0}  \left[ K_n ( \qvec_1 , \dots , \qvec_{p-1}, \qvec_{1;p} - \qvec_{1;p-1} , \kvec_{p+1} , \dots , \kvec_n )  \right]  =  \frac{q_{1;p}^i A^i}{| \qvec_{1;p}|^2} \ , 
\ee
where $A^i$ has units of $k$, and is in general made up of all possible dot products of the wavevector arguments, except for $| \qvec_{1;p}|^2$, which has been factored out.  The fact that a $q^i_{1;p}$ appears in the numerator comes from the fact that the the Laplacian in the denominator always comes with at least one gradient in the numerator, i.e. always at least in the combination $\partial_i / \partial^2$.  Now, we want to take the vector limit $\qvec_{1;p} \rightarrow 0$, so we can Taylor expand $A^i$ as
\be \label{aiexpand}
A^i \approx A^i_{(0)} + q^j_{1;p} A^{ij}_{(1)} + q^j_{1;p} q^k_{1;p} \left( A_{(2)}^i \delta_K^{jk} + A_{(3)}^k \delta_K^{ij} + A^{ijk} \right) + \mathcal{O}(q_{1;p}^3 )  \ ,
\ee
where all of the $A$ coefficients can depend on all of the wavenumber arguments except $\qvec_{1;p}$, and $A^{ijk}$ does not contain any Kronecker delta functions with free indices (since then those terms could be absorbed in $A^i_{(2)}$ and $A^i_{(3)}$).  Since we have already taken the magnitude limit, $A^i_{(2)}=0$ and $A^i_{(3)}=0$, and $A^i_{(0)} = 0$ because there is no pole when $\qvec_{1;p} \rightarrow 0$ (remember that in the kernels the equations of motion have been used already).  Additionally, $A^{ij}_{(1)} = 0$, but this is a bit more non-trivial to see.  In order to have the appropriate form after the magnitude limit in \eqn{maglim1}, which is $q^i_{1;p} q^j_{1;p} / |\vec q_{1;p}|^2$, a term proportional to $A^{ij}_{(1)}$ would have to come from a term like 
\be \label{bcexpand}
\delta^{(n)}_g \supset B^{(n-p)}_{ij} r_{ij}^{(p)} + C_{ij}^{(n-p-m)} \partial_k r_{ij}^{(p)} \frac{\partial_k \theta^{(m)}}{\partial^2} + \dots  \ , 
\ee
where on $B_{ij}$ and $C_{ij}$ the superscript indicates the total perturbative order, and the ellipsis above indicates further terms in the fluid expansion of $r_{ij}$ (which can equally be replaced with $p_{ij}$ above).  If $r_{ij}$ (or equally $p_{ij}$) did not appear at $p$-th order, it would not have the correct form to match the $A^{ij}_{(1)} $ term after the magnitude limit, i.e. the form $q^i_{1;p} q^j_{1;p}A^{ij}_{(1)}  / |\vec q_{1;p}|^2$ in \eqn{maglim1}.  Then, when taking the vector limit $\qvec_{1;p} \rightarrow 0$,  since mass and momentum conservation of dark matter means that $r_{ij}^{(p)} \rightarrow \mathcal{O}(q_{1;p}^2)$, the $B^{(n-p)}_{ij} r_{ij}^{(p)}$ term would contribute a term going like $\mathcal{O}(q_{1;p}^2)$ in \eqn{maglim1}, which is subleading to the $A^{ijk}$ term from \eqn{aiexpand} which goes as $\mathcal{O}(q_{1;p}^1)$ in \eqn{maglim1}.  The $C_{ij}^{(n-p-m)}$ term from \eqn{bcexpand} would contribute at even higher order because there is an extra derivative on $r_{ij}^{(p)}$.  So it must indeed be that $A^{ij}_{(1)} = 0$.  Thus, we see that when we take the vector limit $\qvec_{1;p} \rightarrow 0$, the first term that is non-zero goes like $\sim \mathcal{O}(q_{1;p}^2)$ in $A^i$, which means that \eqn{maglim1} goes like  $\sim \mathcal{O}(q_{1;p}^1)$.  Thus, the final limit that we take that will reproduce the results of \secref{plimsec} is 
\be
 \mathcal{V}^{(1)}_{ \qvec_{1;p} \rightarrow 0}  \left[  \mathcal{M}^{(-2)}_{ |\qvec_{1;p}| \rightarrow 0}  \left[ K_n ( \qvec_1 , \dots , \qvec_{p-1}, \qvec_{1;p} - \qvec_{1;p-1} , \kvec_{p+1} , \dots , \kvec_n )  \right] \right]  =  \frac{q_{1;p}^i  q_{1;p}^j q_{1;p}^k A^{ijk} }{| \qvec_{1;p}|^2}  \ . 
\ee

Now, in \secref{plimsec}, we said that this $p$-particle soft limit is equal to the right-hand side of \eqn{deltagplimmain}, with the total momentum of $\theta^{(p)}  $ going to zero.  However, since in the above manipulations \eqn{maglim1} we also sent the magnitude to zero, we will have to do that on the right-hand side of \eqn{deltagplimmain} as well, and in practice, on $\partial_i \theta^{(p)} / \partial^2$.  So, consider $\partial_i \theta^{(p)} / \partial^2$ in Fourier space
\be \label{velogp}
\frac{q_{1;p}^i }{| \qvec_{1;p}|^2} G_p ( \qvec_1 , \dots , \qvec_{p-1} , \qvec_{1;p} - \qvec_{1;p-1} ) \ . 
\ee
We first must send the magnitude to zero, and then take the vector limit.  Since in \eqn{velogp} there is already a factor of $| \qvec_{1;p}|^{-2}$, we take the zeroth order for the magnitude limit of $G_p$.  Then, because mass and momentum conservation means that $G_p$ has to start at $\sim \mathcal{O}(q_{1;p}^2)$, the leading term is obtained by\footnote{It is interesting to compare this to the normal vector limit of $G_p$, which has to go like $\sim \mathcal{O}(q_{1;p}^2)$ because of mass and momentum conservation
\be
\lim_{\qvec_{1;p} \rightarrow 0} G_p ( \qvec_1 , \dots , \qvec_{p-1} , \qvec_{1;p} - \qvec_{1;p-1} )  = q^i_{1;p} q^j_{1;p} \left( A \delta_K^{ij} + A^{ij} \right) \ ,
\ee
where $A^{ij}$ does not contain any explicit $\delta_K^{ij}$.  Since in \eqn{Gplim} we had to send the magnitude $| \qvec_{1;p}| \rightarrow 0$ first, we see that the $A$ term drops out in \eqn{Gplim}.  The fact that we can take the magnitude limit $| \qvec_{1;p}| \rightarrow 0$ in \eqn{Gplim} and still be left with $q^i_{1;p} q^j_{1;p} A^{ij}$ is essentially what allows us to pick out the appropriate term in the soft limit.  }  
\be  \label{Gplim}
 \mathcal{V}^{(2)}_{ \qvec_{1;p} \rightarrow 0}  \left[  \mathcal{M}^{(0)}_{ |\qvec_{1;p}| \rightarrow 0}  \left[ G_p ( \qvec_1 , \dots , \qvec_{p-1} , \qvec_{1;p} - \qvec_{1;p-1} )  \right]  \right]  =  q_{1;p}^i  q_{1;p}^j  A^{ij}  \ .
\ee
On the full expression \eqn{velogp}, this is equivalent to 
\be
 \mathcal{V}^{(1)}_{ \qvec_{1;p} \rightarrow 0}  \left[  \mathcal{M}^{(-2)}_{ |\qvec_{1;p}| \rightarrow 0}  \left[ \frac{q_{1;p}^i }{| \qvec_{1;p}|^2} G_p ( \qvec_1 , \dots , \qvec_{p-1} , \qvec_{1;p} - \qvec_{1;p-1} )    \right]  \right]  =  q_{1;p}^i  q_{1;p}^j  A^{ij}  \ .
\ee
Restoring the appropriate coefficients in Fourier space, this leads to our final result
\begin{align}
\begin{split} \label{ourppartlim}
& \mathcal{V}^{(1)}_{ \qvec_{1;p} \rightarrow 0}   \left[   \mathcal{M}^{(-2)}_{| \qvec_{1;p}| \rightarrow 0}  \left[ K_n ( \qvec_1 , \dots , \qvec_{p-1}, \qvec_{1;p} - \qvec_{1;p-1} , \kvec_{p+1} , \dots , \kvec_n )   \right] \right] =  \\
& \hspace{1in} \frac{(n-p)! (p-1)!}{n!}   K_{n-p} ( \kvec_{p+1}, \dots , \kvec_n )  \\
&  \hspace{1in} \times  \mathcal{V}^{(1)}_{ \qvec_{1;p} \rightarrow 0}   \left[   \mathcal{M}^{(-2)}_{| \qvec_{1;p}| \rightarrow 0}  \left[ \frac{\qvec_{1;p} \cdot \kvec_{p+1;n}}{|\qvec_{1;p}|^2}   G_p ( \qvec_1 , \dots , \qvec_{n-1},  \qvec_{1;p} - \qvec_{1;p-1} ) \right] \right] \ ,
\end{split}
\end{align}
for $p \geq 2$, while the $p = 1$ case is covered by \eqn{k1kerlims}.  We have explicitly verified \eqn{ourppartlim} up to $n = 6$ for $p = 2, \dots , n-1$.   Although all of the kernels that we presented in this paper already must satisfy \eqn{ourppartlim} by construction, it serves as a highly non-trivial cross check of our results.  We also see that \eqn{ourppartlim} is quite similar to the analogous expression in \cite{DAmico:2021rdb}, but we have now given an explicit prescription for how to verify it directly given a set of kernels $K_n$ as a function of the external momenta.

\section*{Acknowledgments}

We thank the organizers and participants of the ``Galaxies meet QCD'' conference where part of this work was initiated.  A.E. thanks John Joseph Carrasco, Cong Shen, Sasank Chava, and Sissi Chen for discussions on this and related projects.  This work was supported by a grant from the Simons Foundation International [SFI-MPS-SSRFA-00012751, AE].  A.E. is also supported in part by the DOE under contract DE-SC0015910, and by Northwestern University via the Amplitudes and Insight Group, Department of Physics and Astronomy, and Weinberg College of Arts and Sciences.  L.S. is supported by the SNSF grant 200021 213120.

%
%

\begin{appendix}

%
%
\section{Dark-matter solution} \label{dmsolapp}

The equations of motion for the long-wavelength dark-matter fields are given by  \cite{Baumann:2010tm, Carrasco:2012cv}
\begin{equation}
\label{eomvel3}
\begin{split}
& a H \delta'   +  a^{-1} \partial_i((1 + \delta) v^i) = 0 \ ,  \\
&a H v^i{}' + H   v^i  + a^{-1}\partial_i \Phi    + a^{-1} v^j \partial_j v^i    = - a^{-1} \frac{\partial_j \tau^{ij} }{\bar \rho ( 1 + \delta )} \ ,
\end{split}
\end{equation}
as well as the Poisson equation
\be \label{poisson}
a^{-2} \partial^2 \Phi = \frac{3}{2} \om H^2 \delta  \ ,
\ee
where $\tau^{ij}$ is a local-in-space and non-local-in-time \cite{Carrasco:2013mua, Carroll:2013oxa} function of the long-wavelength fields and stochastic fields.  The solution of these equations up to fifth order, including all relevant non-local-in-time counterterms, is given in \cite{Anastasiou:2025jsy}, where the two-loop power spectrum was also computed.  Since we focus on the bias expansion in this work, that is at lowest order in spatial derivatives, and since the dark-matter counterterms enter with at least two derivatives, we can take $\tau^{ij} = 0$ for the rest of this work.  Inclusion of higher-derivative bias terms is a straightforward extension of this work (following the non-local-in-time expansion of $\tau^{ij}$ in \cite{Anastasiou:2025jsy}, for example) which we leave for future study.  

We perturbatively expand the dark-matter overdensity and velocity field, which can be written as
\be
\delta(\vec k,a)=\sum_n \delta^{(n)}(\vec k,a)  \andd  v^i ( \kvec , a ) = \sum_n v^i_{(n)} ( \kvec , a) \ ,    
\ee
where $\delta^{(n)}(\vec k,a)$ is of order $[\delta^{(1)}(\vec k,a)]^n$ and $\delta^{(1)}(\vec k,a)$ is the solution to the linearized equations. 
The solution to the linear equation, $ \delta^{(1)}(\vec k,a)$, factorizes into a time dependent and a space dependent part
\be
\delta^{(1)}(\vec k,a)=D(a)\tilde\delta^{(1)}_{\kvec} \ ,
\ee
where $D(a)$ is known as the growth factor and satisfies
\be \label{daeqn}
 a^2 D''(a) + \left(2 + \frac{a \mathcal{H}'}{\mathcal{H}} \right)aD'(a) - \frac{3}{2}\Omega_m D(a)=0\ ,
\ee 
where $\cH \equiv a H$, while $\tilde\delta^{(1)}_{\kvec}$ is fixed by the initial conditions at some time $a_{\rm in}$.\footnote{We normalize the growth factor  such that $D(a_{\rm in}) = 1$ which implies that $\delta^{(1)}(\vec k,a_{\rm in})=\tilde\delta^{(1)}_{\kvec} $.  We also assume Gaussian initial conditions, and the initial linear power spectrum $P_{11} ( k)$ is defined by
\be
\langle \tilde \delta^{(1)}_{\kvec } \tilde \delta^{(1)}_{\kvec'} \rangle = ( 2 \pi)^3 \delta_D ( \kvec + \kvec') P_{11} ( k )  \ . 
\ee}
In the so-called EDS approximation, which works at per cent level and even better at higher redshifts  (see for example~\cite{Donath:2020abv,Fasiello:2022lff,Garny:2022fsh}), we have $\Omega_m(a) \approx f(a)^2$, with 
$f (a) \equiv \frac{a D'(a)}{D(a)}$, and the $n$-th order solution  $\delta^{(n)}(\vec k,a)$ has a factorized time dependence\footnote{The time dependence factorizes from the space dependence also with exact time dependence (see \cite{Lewandowski:2017kesFF}, for example), so the procedure that we use in this work still applies in that case.  However, there can be more bias parameters \cite{Desjacques:2016bnm}, although they are expected to be suppressed by how close the exact time dependence is to the EdS approximation. } 
\be \label{edsdelta}
\delta^{(n)}(\vec k,a)=D(a)^n \tilde\delta^{(n)}(\vec k)\ ,
\ee 
such that $\tilde \delta^{(n)}$ is time-independent.  The velocity field, neglecting vorticity,\footnote{ For our purposes, this is a good assumption, see \appref{vorticityapp}.} at a given order can be rewritten in terms of its rescaled divergence  $\theta$, defined by 
\be \label{thetadefa}
\theta(\vx, t') \equiv  -  \frac{1}{a H  f }\partial_i v^i(\vx, t')  \ , 
\ee
as
\begin{align} \label{vastheta}
    v^{i}_{(n)} (\vx, t') = -a H f\frac{\partial_i}{\partial^2}\theta^{(n)}(\vx, t') = -\frac{a \dot{D}(t')}{D(t')}\frac{D(t')^n}{D(t)^n}\frac{\partial_i}{\partial^2}\theta^{(n)}(\vx, t)\, ,
\end{align}
This means that we can also define the time-independent velocity divergence $\tilde \theta$ from
\be
\theta^{(n)} ( \kvec , a ) = D(a)^{n} \tilde \theta^{(n)} ( \kvec)  \ . 
\ee
The velocity divergence is rescaled in \eqn{thetadefa} such that $\theta^{(1)} ( \kvec , a ) = \delta^{(1)} ( \kvec , a )$.

The solutions to the time-independent density and velocity fields, assuming $\tau^{ij} =0$, can be written in the form,
\begin{align}
\label{eq:SPTdellta}
    \tilde{\delta}^{(n)}(\vec{k}) &= \int_{\qvec_1 , \dots , \qvec_n}^{\kvec} F_n(\vec{q}_1,\dots,\vec{q}_{n})\tilde{\delta}^{(1)}_{\qvec_1} \dots\tilde{\delta}^{(1)}_{\qvec_n} \, ,\\
    \label{eq:SPTvelo}
    \tilde \theta^{(n)} (\vec{k}) &= \int_{\qvec_1 , \dots , \qvec_n}^{\kvec} G_n(\vec{q}_1,\dots,\vec{q}_{n})\tilde{\delta}^{(1)}_{\qvec_1} \dots\tilde{\delta}^{(1)}_{\qvec_n} \, ,
\end{align}
and assuming that the velocity field is irrotational, it is given by
\be
v^{i}_{(n)} ( \kvec , a ) = i a H f D(a)^n \frac{k^i}{k^2} \tilde \theta^{(n)} ( \kvec) \ .
\ee
Above, $F_n$ and $G_n$ are the symmetric kernels describing the dark-matter solution (in general, all perturbative kernels in this paper are assumed to be symmetric, unless otherwise stated).  In terms of the dark-matter interaction vertices $\alpha$ and $\beta$,
\begin{align}  
\alpha ( \qvec_1 , \qvec_2 ) = 1 + \frac{\qvec_1 \cdot \qvec_2}{q_1^2}  \andd
\beta( \qvec_1 , \qvec_2 )  = \frac{ | \qvec_1 + \qvec_2 |^2 \qvec_1 \cdot \qvec_2}{2 q_1^2 q_2^2}    \label{betadef2new} \ ,
 \end{align}
$F_n$ and $G_n$ are explicitly given by \cite{Goroff:1986ep,Jain:1993jh}
\begin{align}
\begin{split}
& F_n ( \kvec_1 , \dots , \kvec_n) = \underset{\kvec_1, \dots , \kvec_n}{\text{sym.}} \sum_{m=1}^{n-1} \frac{G_m ( \kvec_1 , \dots , \kvec_m) }{(2 n + 3)(n-1)} \Bigg( (1+ 2 n) \alpha ( \kvec_{1;m} , \kvec_{m+1;n}  ) F_{n-m} ( \kvec_{m+1} , \dots , \kvec_n  )   \\
& \hspace{2.5in} + 2 \beta ( \kvec_{1;m} , \kvec_{m+1;n}  ) G_{n-m} ( \kvec_{m+1} , \dots , \kvec_n  )    \Bigg) \ , \\ 
& G_n ( \kvec_1 , \dots , \kvec_n) =  \underset{\kvec_1, \dots , \kvec_n}{\text{sym.}} \sum_{m=1}^{n-1} \frac{G_m ( \kvec_1 , \dots , \kvec_m) }{(2 n + 3)(n-1)} \Bigg( 3 \alpha ( \kvec_{1;m} , \kvec_{m+1;n}  ) F_{n-m} ( \kvec_{m+1} , \dots , \kvec_n  )   \\
& \hspace{2.5in} + 2 n  \beta ( \kvec_{1;m} , \kvec_{m+1;n}  ) G_{n-m} ( \kvec_{m+1} , \dots , \kvec_n  )    \Bigg)  \ , 
\end{split}
\end{align}
where we have defined $\kvec_{i;j} \equiv \kvec_i + \dots + \kvec_j$,  $F_1 = G_1 = 1$, and
\be
\underset{\kvec_1, \dots , \kvec_n}{\text{sym.}} f(\kvec_1 , \dots , \kvec_n)  \equiv \frac{1}{n!} \sum_{\sigma \in S_n} f ( \sigma ( \kvec_1 , \dots , \kvec_n))  \ ,
\ee
where $S_n$ is the permutation group of $n$ elements.

Finally, this means that we can write the building blocks of the bias expansion as
\be \label{rijpert}
r^{ij}_{(n)} ( \kvec , t' ) = \left( \frac{D(t')}{D(t)} \right)^n r^{ij}_{(n)} ( \kvec , t ) = D(t')^n \frac{k^i k^j}{k^2} \tilde \delta^{(n)} ( \kvec )  \ ,
\ee
and
\be \label{pijpert}
p^{ij}_{(n)} ( \kvec , t' ) = \left( \frac{D(t')}{D(t)} \right)^n p^{ij}_{(n)} ( \kvec , t ) = D(t')^n \frac{k^i k^j}{k^2} \tilde \theta^{(n)} ( \kvec )  \ . 
\ee
 Notice that the $p^{ij}$ above is a symmetric tensor because we can neglect vorticity for the bias expansion, i.e. at lowest order in derivatives, see \appref{vorticityapp}.

%
\section{Dark-matter velocity vorticity} \label{vorticityapp}

In this appendix, we show that for the bias expansion that we do in this work, which is at zeroth order in derivatives and for non-stochastic fields, we can ignore the dark-matter velocity vorticity.   In terms of the rescaled velocity divergence $\theta = - \partial_i v^i / (a H f)$ and the rescaled velocity vorticity 
\be
\omega^i = \frac{-\epsilon^{ijk} \partial_j v^k}{a H f} \ ,  
\ee
we can decompose the velocity field as
\be
v^i = - a H f \left( \frac{\partial_i \theta }{\partial^2} - \epsilon^{ijk} \frac{\partial_j \omega^k}{\partial^2} \right) \ . 
\ee
 In these variables, the dark-matter equations of motion \eqn{eomvel3}, including vorticity, become \cite{Carrasco:2013mua}
\begin{align}
\begin{split} \label{vorticityeom}
 \frac{a}{f} \delta '  - \theta & = \delta \theta + \partial_i \delta \frac{\partial_i \theta}{\partial^2} -  \epsilon^{ijk} \partial_i \delta \frac{\partial_j \omega^k}{\partial^2}  \ ,  \\
 \frac{a}{f} \theta ' - \theta + \frac{3 \Om}{2 f^2} \left( \theta - \delta \right)  & = \partial_i \theta  \left( \frac{\partial_i \theta }{\partial^2} - \epsilon^{ijk} \frac{\partial_j \omega^k}{\partial^2} \right) \\
 &\quad  +\left( \frac{\partial_i \partial_j \theta }{\partial^2} - \epsilon^{ikl} \frac{\partial_k \partial_j \omega^l}{\partial^2} \right)\left( \frac{\partial_i \partial_j \theta }{\partial^2} - \epsilon^{jkl} \frac{\partial_k \partial_i \omega^l}{\partial^2} \right) \\
 & \quad  + \frac{1}{a^2 H^2 f^2} \partial_i \left( \frac{\partial_j \tau^{ij} }{\bar \rho ( 1 + \delta )}   \right) \ , \\
 \frac{a}{f} \omega^i{}' + \left( \frac{3 \Om}{2 f^2} - 1 \right) \omega^i  & = -\epsilon^{ijk} \partial_j \left( \epsilon^{klm} \left( \frac{\partial_l \theta }{\partial^2} - \epsilon^{lnp} \frac{\partial_n \omega^p}{\partial^2} \right) \omega^m - \frac{\partial_l \tau^{kl}}{a^2 H^2 f^2 \bar \rho ( 1 + \delta) } \right)  \ ,
\end{split}
\end{align}
where in the last equation we used that\footnote{As a reminder, $\epsilon^{ijk} \epsilon^{abk} = \delta^{ia}\delta^{jb}  - \delta^{ib} \delta^{ja} $.}
\be
 \frac{1}{a^2 H^2 f^2} \epsilon^{ijk} \partial_j ( v^l \partial_l v^k ) =  \frac{1}{a H f} \epsilon^{ijk} \partial_j \left( \epsilon^{klm} v^l \omega^m \right)  = - \epsilon^{ijk} \partial_j \left(\epsilon^{klm}  \left( \frac{\partial_l \theta }{\partial^2} - \epsilon^{lnp} \frac{\partial_n \omega^p}{\partial^2} \right) \omega^m \right)  \ . 
\ee

Thus, the linear equation for the rescaled velocity vorticity is
\be
\frac{a}{f} \omega^i_{(1)}{}' +  \left( \frac{3 \Om}{2 f^2} - 1 \right) \omega^i_{(1)} = 0 \ ,   
\ee
which is solved by
\be
\omega^i_{(1)} ( \xvec , a ) = \frac{1}{a^2 H(a) f(a)} \tilde \omega^i_{(1)} (\xvec ) \ . 
\ee
In this EdS approximation, this scales as $a^{-1/2}$, which should be compared to $\theta^{(1)}$, which scales as $a^1$.  Thus, any vorticity in the initial conditions will decay with respect to the growing modes like $a^{-3/2}$, so we can take $\omega^i_{(1)} ( \xvec , a ) = 0$ to great accuracy.  Even more, neglecting the stress tensor, the second order vorticity is sourced non-linearly by a term that is proportional to the vorticity itself
\be
\frac{a}{f} \omega^i_{(2)}{}' +  \left( \frac{3 \Om}{2 f^2} - 1 \right) \omega^i_{(2)}  = -\epsilon^{ijk} \partial_j \left( \epsilon^{klm} \left( \frac{\partial_l \theta^{(1)} }{\partial^2} - \epsilon^{lnp} \frac{\partial_n \omega^p_{(1)}}{\partial^2} \right) \omega^m_{(1)} \right)  \ ,
\ee
so that $\omega^i_{(2)}$ is also zero.  This obviously continues to all higher orders if the stress tensor is ignored.\footnote{This is the familiar statement that vorticity is not generated in standard perturbation theory.}  However, as pointed out in \cite{Carrasco:2013mua}, and as can be seen from \eqn{vorticityeom}, vorticity is generated by the stress tensor $\tau^{ij}$ in the EFT of LSS.  Indeed, it was shown in \cite{DAmico:2022ukl} that in order to renormalize the one-loop bispectrum, even for dark matter without redshift space distortions, vorticity \emph{needs} to be generated.  Although it is generated at second order, it does not contribute to $\delta$ or $\theta$ until third order, as can be seen in \eqn{vorticityeom}.  In \cite{Anastasiou:2025jsy} it was shown that one must include this contribution from the vorticity to consistently renormalize the two-loop dark-matter power spectrum.  See also~\cite{Garny:2022kbk} for a study of vorticity in perturbation theory.

Since we do not compute the stochastic contributions, we focus on the response counterterms in the stress tensor.  Looking at the vorticity equation in \eqn{vorticityeom},  we see that the contribution from the first-order stress tensor is zero since $\tau^{kl}_{(1)} \sim c_1 \delta^{kl}_K \partial^2 \Phi^{(1)} + c_2 \partial_k \partial_l \Phi^{(1)}$ and $\epsilon^{ijk} \partial_j  \partial_l \left( c_1 \delta^{kl}_K \partial^2 \Phi^{(1)} + c_2 \partial_k \partial_l \Phi^{(1)} \right) = 0$.  Next we look at the second-order terms.  The terms proportional to $\omega^m$ are zero since $\omega_{(1)}^m = 0$.  One possible second-order contribution comes from expanding $1/(1 + \delta)$ and is proportional to
\begin{align}
\begin{split}
- \epsilon^{ijk} \partial_j ( \delta^{(1)} \partial_l \tau^{kl}_{(1)} )  & \propto - \epsilon^{ijk} \partial_j ( \partial^2 \Phi^{(1)} \partial_l (c_1 \delta^{kl}_K \partial^2 \Phi^{(1)} + c_2 \partial_k \partial_l \Phi^{(1)}) )  \\
& = -  ( c_1 + c_2 ) \epsilon^{ijk} \partial_j ( \partial^2 \Phi^{(1)} \partial_k\partial^2 \Phi^{(1)} ) \\
&=  -  \frac{( c_1 + c_2 )}{2} \epsilon^{ijk} \partial_j \partial_k( \partial^2 \Phi^{(1)} \partial^2 \Phi^{(1)} ) \ , 
\end{split}
\end{align}
which is zero.  So, the first non-zero contribution comes from the second-order stress tensor which in the EdS approximation can be written as
\be
\tau^{ij}_{(2)} ( \xvec , a )  = \frac{f^2 H^2 \bar \rho}{\knl^2} D(a)^4 \tilde \tau^{ij}_{(2)}  ( \xvec) \ , 
\ee
so that in the EdS approximation, writing $\omega^i_{(2)} ( \xvec , a ) = D(a)^4 \tilde \omega^i_{(2)} (\xvec)$, we have
\be
\tilde \omega^i_{(2)} ( \xvec ) = \frac{2}{9 \knl^2} \epsilon^{ijk} \partial_j \partial_l \tilde \tau^{kl}_{(2)} ( \xvec )  \ . 
\ee
Thus, the vorticity is a higher-derivative term, suppressed by $\knl^2$.  The fact that the vorticity is generated with at least two spatial derivatives continues to higher orders, as can be readily seen from the last equation in \eqn{vorticityeom}.  Now, the velocity building block of the bias expansion is
\be
p^{ij} =   \frac{\partial_i \partial_j \theta }{\partial^2} - \epsilon^{jkl} \frac{\partial_i \partial_k \omega^l}{\partial^2}  \ ,
\ee
but since we only consider the bias expansion at zeroth order in spatial derivatives in this work, we see that we can ignore the vorticity for that purpose, in which case $p^{ij}$ is symmetric.  However, when considering higher-derivative bias, these terms may need to be included.

%
%
\section{Details for perturbative expansion} \label{pertexpapp}

Here we detail the steps taken to derive our final expression for the perturbative expansion of the galaxy overdensity \eqn{finalfulltauij} in terms of the new kernels \eqn{finalcndef} derived in this work.
We start with \eqn{oxflprodexp} and expand up to $n$-th order in perturbations
\begin{align}
\begin{split}
 \left[ \mathcal{O}_{m} \big|_{\xfl}  ( \xvec , t ; t_1 , \dots , t_m )  \right]^{(n)}   & = \Delta_{\mathcal{O}_m} [ i_1 , j_1 , \dots , i_m , j_m ]   \sum_{\{ n_1 , \dots , n_m \}_n}  \\
 & \times  [ t^{i_1 j_1}_{\mathcal{O}_m , 1} ( \xfl ( \xvec , t , t_1) , t_1 ) ]^{(n_1)}  \cdots [ t^{i_m j_m}_{\mathcal{O}_m , m} ( \xfl ( \xvec , t , t_m)  , t_m ) ]^{(n_m)}
\end{split}
\end{align}
where $\{ n_1 , \dots , n_m \}_n$ is the set of all $(n_1 , \dots , n_m)$ such that $n_1 + \cdots + n_m = n$ with $n_a \geq 1$ for $a = 1 , \dots , m$.  

We further separately expand the fluid trajectory at each time using the expansion of \cite{Donath:2023sav}, 
\be \label{texpand}
 [ t_{\mathcal{O}_m,a}^{i_a j_a}( \xfl ( \xvec , t , t_a)  , t_a ) ]^{(n_a)} =  \sum_{\alpha_a = 1}^{n_a} \left( \frac{D(t_a)}{D(t)} \right)^{\alpha_a}  [ \mathbb{C}_{ (\mathcal{O}_m , a)  , \alpha_a}^{ i_a j_a}]^{(n_a)} ( \xvec , t )  \ ,
\ee
which is a special case with $m = 1$ of the general formula, since each of the $t_{\mathcal{O}_m,a}$ is either $r$ or $p$.  As derived in \cite{Donath:2023sav}, they satisfy the following recursion relation
\be \label{singlerec}
[\mathbb{C}^{ij}_{r, \alpha} ]^{(n)} ( \xvec , t ) = \frac{1}{n - \alpha } \sum_{\ell = \alpha}^{n-1} \partial_k [\mathbb{C}^{ij}_{r, \alpha} ]^{(\ell)} ( \xvec , t )  \frac{\partial_k}{\partial^2} \theta^{( n - \ell) } ( \xvec , t )  \ , 
\ee
for $\alpha < n$, along with the equal-time completeness relation to determine the $ \alpha = n$ kernel
\be \label{singleet}
r^{ij}_{(n)}( \xvec , t ) = \sum_{\alpha = 1}^{n} [\mathbb{C}^{ij}_{r, \alpha} ]^{(n)} ( \xvec , t )  \ , 
\ee
and analogously for $p^{ij}$ (explicit expressions for $r^{ij}_{(n)}$ and $p^{ij}_{(n)}$ are given in \eqn{rijpert} and \eqn{pijpert} respectively).  Thus, these functions can be easily generated using the methods of \cite{Donath:2023sav}.

Plugging \eqn{texpand} into \eqn{oxflprodexp} and then into \eqn{generaltauij}, we obtain 
\begin{align}
\begin{split}
\delta_g^{(n)} ( \xvec , t ) & = \sum_{m = 1}^n \sum_{\{\mathcal{O}_m\}} \int^t \, d t_1 \cdots d t_m H ( t_1 ) \cdots H(t_m )  \, c_{\mathcal{O}_m } ( t , t_1 , \dots , t_m )   \Delta_{\mathcal{O}_m} [ i_1 , j_1 , \dots , i_m , j_m ]   \\
& \times \sum_{\{ n_1 , \dots , n_m \}_n}  \prod_{a = 1}^m \left[ \sum_{\alpha_a = 1}^{n_a} \left( \frac{D(t_a)}{D(t)} \right)^{\alpha_a}  [ \mathbb{C}_{ (\mathcal{O}_m , a)  , \alpha_a}^{ i_a j_a}]^{(n_a)} ( \xvec , t )   \right]  \ ,
\end{split}
\end{align}
which, by exchanging the sum and product in the second line, we can rewrite it as 
\begin{align}
\begin{split} \label{timeints}
\delta_g^{(n)} ( \xvec , t ) & = \sum_{m = 1}^n \sum_{\{\mathcal{O}_m\}} \int^t \, d t_1 \cdots d t_m H ( t_1 ) \cdots H(t_m )  \, c_{\mathcal{O}_m } ( t , t_1 , \dots , t_m )   \Delta_{\mathcal{O}_m} [ i_1 , j_1 , \dots , i_m , j_m ]   \\
& \times \sum_{\{ n_1 , \dots , n_m \}_n} \sum_{\alpha_1 = 1}^{n_1} \dots  \sum_{\alpha_m = 1}^{n_m}  \left[  \prod_{a = 1}^m  \left( \frac{D(t_a)}{D(t)} \right)^{\alpha_a}  [ \mathbb{C}_{ (\mathcal{O}_m , a)  , \alpha_a}^{ i_a j_a}]^{(n_a)} ( \xvec , t )   \right]  \ .
\end{split}
\end{align}
Next, we define EFT coefficients, which are only a function of the final time $t$, as integrals over the original time-dependent functions $c_{\mathcal{O}_m} ( t , t_1 , \dots , t_m)$ and factors of $D$, 
\begin{align}
\begin{split} \label{gentimefns}
c_{\mathcal{O}_m; \alpha_1 , \dots , \alpha_m } (t ) \equiv  \int^t \, d t_1 \cdots d t_m H ( t_1 ) \cdots H(t_m )  \, c_{\mathcal{O}_m } ( t , t_1 , \dots , t_m )    \prod_{a = 1}^m  \left( \frac{D(t_a)}{D(t)} \right)^{\alpha_a}  \ , 
\end{split}
\end{align}
so that \eqn{timeints} becomes 
\begin{align}
\begin{split}  \label{intereq}
\delta_g^{(n)} ( \xvec , t ) & = \sum_{m = 1}^n \sum_{\{\mathcal{O}_m\}}   \sum_{\{ n_1 , \dots , n_m \}_n} \sum_{\alpha_1 = 1}^{n_1} \dots  \sum_{\alpha_m = 1}^{n_m}    \\
& \times c_{\mathcal{O}_m; \alpha_1 , \dots , \alpha_m } (t )  \left(  \Delta_{\mathcal{O}_m} [ i_1 , j_1 , \dots , i_m , j_m ]   \prod_{a = 1}^m   [ \mathbb{C}_{ (\mathcal{O}_m , a)  , \alpha_a}^{ i_a j_a}]^{(n_a)} ( \xvec , t )    \right) \ .
\end{split}
\end{align}

The time-dependent functions $c_{\mathcal{O}_m; \alpha_1 , \dots , \alpha_m } (t ) $ are free EFT coefficients, so they determine the separate functional forms that can appear.  To get to the final form \eqn{finalfulltauij}, we need to swap the sum over the $n_a$ with the sum over the $\alpha_a$ using   
\be
  \sum_{\{ n_1 , \dots , n_m \}_n} \sum_{\alpha_1 = 1}^{n_1} \dots  \sum_{\alpha_m = 1}^{n_m}  =  \sum_{\{\alpha_1, \dots , \alpha_m\}_{\leq n} } \sum_{\{n_1 , \dots , n_m\}^\alpha_n }
\ee
where $\{\alpha_1, \dots , \alpha_m\}_{\leq n} $ is the set of all $(\alpha_1 , \dots , \alpha_m)$ with $\alpha_1 + \cdots + \alpha_m \leq n$ and $\alpha_a \geq 1$, and $\{n_1 , \dots , n_m\}^\alpha_n $ is the set of all $(n_1 , \dots , n_m )$ such that $n_1 + \cdots + n_m = n$ and $n_a \geq \alpha_a$ for $a = 1 , \dots , m$.  

As a further remark, the local-in-time expansion is obtained by taking
\be \label{litcoeffdeltafn}
c_{\mathcal{O}_m } ( t , t_1 , \dots , t_m ) = c_{\mathcal{O}_m}(t) \frac{\delta_D(t - t_1)}{H(t_1)} \cdots \frac{\delta_D(t - t_m)}{H(t_m)} \ ,
\ee
in \eqn{generaltauij}, i.e. the response is purely local in time. On the other hand, as used in \cite{Donath:2023sav}, the case when all operators are inserted at the same fluid time is obtained by taking 
\be \label{singletimelimit}
c_{\mathcal{O}_m} ( t , t_1 , \dots , t_m ) =  c_{\mathcal{O}_m} ( t , t_1 ) \frac{\delta_D ( t _1 - t_2)}{H(t_2)} \frac{\delta_D ( t _1 - t_3)}{H(t_3)} \dots \frac{\delta_D ( t _1 - t_m)}{H(t_m)}  \ ,
\ee
in \eqn{generaltauij}, which leads to all fields being inserted at and integrated over the same time $t_1$.

Finally, we can derive another set of equations that relates our kernels to the kernels of \cite{Donath:2023sav} by taking $t_1 = \cdots = t_m = t'$ in \eqn{fulloxflexp}.  This gives
\begin{align}
\begin{split}
 \left[ \mathcal{O}_{m} \big|_{\xfl}  ( \xvec , t ; t' , \dots , t' )  \right]^{(n)}  =  [ \mathcal{O}_m ( \xfl ( \xvec , t , t' ) , t' ) ]^{(n)} = \sum_{\alpha=1}^{n - m +1} \left( \frac{D(t')}{D(t)} \right)^{\alpha + m -1} \mathbb{C}_{\mathcal{O}_m , \alpha}^{(n)} ( \xvec , t)  \ .
\end{split}
\end{align}
Since this has to be equal to the right-hand side of \eqn{fulloxflexp} for all $t'$, this gives
\be \label{heretobareq}
 \sum_{\{\alpha_1, \dots , \alpha_m\}_{\alpha} }     \bar{\mathbb{C}}^{(n)}_{\mathcal{O}_m; \alpha_1 , \dots , \alpha_m} ( \xvec , t)  = \mathbb{C}_{\mathcal{O}_m , \alpha-m+1}^{(n)} ( \xvec , t )  \ ,
\ee
for all $m \leq \alpha \leq n$.  This says that the sum over all of the multi-insertion kernels with a fixed total $\sum_{i=1}^m \alpha_i$ is equal to a specific kernel from \cite{Donath:2023sav}.  However, all of these equations with $m \leq \alpha \leq n-1$ are automatically satisfied once the recursion \eqn{newxflrec} is imposed, and so the new equation is 
\be
 \sum_{\{\alpha_1, \dots , \alpha_m\}_{n} }    \bar{\mathbb{C}}^{(n)}_{\mathcal{O}_m; \alpha_1 , \dots , \alpha_m} ( \xvec , t)  = \mathbb{C}_{\mathcal{O}_m , n-m+1}^{(n)} ( \xvec , t )  \ .
\ee

%
%
\section{Proof of $p$-particle soft limits} \label{plimapp}

%
\subsection{Dark-matter $p$-particle soft limits} \label{dmplimapp}
In this section, we will show that the dark-matter equations of motion imply 
\begin{align}
\begin{split} \label{dmirlimits}
& \delta^{(n)} \underset{p}{\rightarrow} \frac{1}{p} \partial_i \delta^{(n-p)} \frac{\partial_i \theta^{(p)} }{\partial^2} \ , \\
& \theta^{(n)} \underset{p}{\rightarrow} \frac{1}{p} \partial_i \theta^{(n-p)} \frac{\partial_i \theta^{(p)} }{\partial^2} \ ,
\end{split}
\end{align}
for $ 1 \leq p \leq n-1$ in the $p$-particle soft limit. In the EdS approximation, the dark-matter equations of motion \eqn{eomvel3} with $\tau^{ij} = 0$ are
\begin{align}
\begin{split} \label{dmeomanbn}
& n \delta^{(n)} - \theta^{(n)} = A_n\ , \\
& n \theta^{(n)} + \frac{1}{2} \theta^{(n)} - \frac{3}{2} \delta^{(n)} = B_n \  ,
\end{split}
\end{align}
where 
\begin{align}
\begin{split} \label{anbndef}
& A_n = \sum_{\ell = 1}^{n-1}  \left(  \partial_i \delta^{(\ell)} \frac{\partial_i \theta^{(n-\ell)}}{\partial^2} + \theta^{(\ell)} \delta^{(n-\ell)}   \right) \ , \\
& B_n = \sum_{\ell = 1}^{n-1}   \left( \partial_i \theta^{(\ell)} \frac{\partial_i \theta^{(n-\ell)}}{\partial^2}  + \frac{\partial_i \partial_j \theta^{(\ell)} }{\partial^2} \frac{\partial_i \partial_j \theta^{(n-\ell)} }{\partial^2} \right)  \ .
\end{split}
\end{align}
and solving \eqn{dmeomanbn} for $\delta^{(n)}$ and $\theta^{(n)}$, we obtain
\begin{align}
\begin{split} \label{deltathetaab}
& \delta^{(n)} = \frac{1}{(2 n +3)(n-1)}  \left[ (1 + 2n) A_n + 2 B_n   \right]  \ , \\
& \theta^{(n)} = \frac{1}{(2 n +3)(n-1)}  \left[ 3  A_n + 2 n B_n   \right]  \ . 
\end{split}
\end{align}

We are going to use induction a number of times to prove the $p$-particle soft limits, so we first establish a few general results.  First, using the solutions \eqn{deltathetaab}, we notice that 
\be
\delta^{(n)}  \underset{n}{\rightarrow} 0  \andd \theta^{(n)}  \underset{n}{\rightarrow} 0  \ , 
\ee
simply because $\partial_i \theta^{(n)} / \partial^2$ is absent.  Now, assume we have two scalars $X$ and $Y$ (like $\delta$ and $\theta$, for example) that satisfy the $p$-particle soft limits 
\begin{align}
\begin{split} \label{xyirlimits}
& X^{(n)} \underset{p}{\rightarrow} \frac{1}{p} \partial_i X^{(n-p)} \frac{\partial_i \theta^{(p)} }{\partial^2} \ , \\
& Y^{(n)} \underset{p}{\rightarrow} \frac{1}{p} \partial_i Y^{(n-p)} \frac{\partial_i \theta^{(p)} }{\partial^2} \ .
\end{split}
\end{align}
for an $n$-th order field, for all $p$ from $n -1$ down to some $n - x$ (with $x>1$), for arbitrary $n$.  We will be interested to show that this implies that \eqn{xyirlimits} is also satisfied for $p = n-x-1$, thus completing the inductive step.  Before getting into the full proof, we can see why induction is an effective way to proceed.  We assume that the $p$-particle limits work on an $n$-th order field for $n-x \leq p \leq n-1$ for all $n$.  Then we want to take the $p = n - x -1$ particle limit of the expression of interest at $n$-th order.  This expression will always be made of products of fields of order $n' \leq n-1$, so we will be taking the $p = n - x -1$ limit of fields at $n'$-th order.  But we will be able to use the induction assumption in these cases because our assumption that the limits work for $p$ in $n-x \leq p \leq n-1 $ for all $n$ means that it works for $p$ in $ n' -x \leq p \leq n' -1$ and now $n' - x \leq n-x-1$.  Thus, we will be able to use the induction assumption for the $p = n - x -1$ particle limit on all of the individual fields that make up the $n$-th order field of interest.

Along the way, we will encounter products of fields, as in \eqn{anbndef}, and we will want to take the $p$-particle soft limit of those.  First, consider the product of scalars
\be \label{xyn}
[XY]^{(n)} = \sum_{\ell=1}^{n-1} X^{(\ell)} Y^{(n-\ell)} \ . 
\ee
For products of fields, we have
\be
X^{(\ell)} Y^{(n - \ell)}  \underset{p}{\rightarrow} X^{(\ell)} \Big|_p Y^{(n - \ell)} + X^{(\ell)} Y^{(n-\ell)} \Big|_p   \ , 
\ee 
where we use the notation 
\be
X^{(\ell)}\underset{p}{\rightarrow}   X^{(\ell)} \Big|_p \ , 
\ee
 to mean the $p$-particle limit of $X^{(\ell)}$, 
since the factor of $\partial_i \theta^{(p)} / \partial^2$ can come from either $X$ or $Y$.  Now, take the $p$-particle soft limit of \eqn{xyn} for $p = n - x -1$, and use the fact that fields can only have non-vanishing $p$-particle soft limits if the order of the field is greater or equal to $p+1$ (we showed this above for $\delta$ and $\theta$, but it will also be true for other relevant fields, which we will discuss below), so we have
\be \label{step1}
[XY]^{(n)}  \underset{p}{\rightarrow} \sum_{\ell = p+1}^{n-1} X^{(\ell)} \Big|_p Y^{(n-\ell)}  + \sum_{\ell = 1}^{n-p-1} X^{(\ell)} Y^{(n-\ell)} \Big|_p \ . 
\ee
In the first sum above, we take the terms with $p$-particle soft limits in $X$, and in the second sum, we take the terms with $p$-particle soft limits in $Y$.  For the $X^{(\ell)}$, we have
\be
X^{(\ell)} \underset{p}{\rightarrow} \frac{1}{p} \partial_i X^{(\ell - p)} \frac{\partial_i \theta^{(p)}}{\partial^2} \ ,
\ee
by the induction assumption \eqn{xyirlimits}.  To see that we can use the induction assumption, notice that for $X^{(\ell)}$, the induction assumption covers $\ell - x \leq p \leq \ell -1$, and the sum in \eqn{step1} involves $X^{(\ell)}$ for values of $\ell $ in $p + 1 \leq \ell \leq n-1$.  Taking the term at the upper limit, $\ell = n-1$, the induction assumption covers $n - x -1 \leq p \leq n-2$, so we see that the value $p = n - x -1$ is within this range, so we can use the induction assumption.  Similarly, at the lower limit $\ell = p+1$, the induction assumption covers $p +1 - x \leq p \leq p$, and the value $p = n - x -1$ is within this range, so we can use the induction assumption again.  The same applies for all other values of $\ell$ in $p + 1 \leq \ell \leq n-1$.  Similarly, we can use the induction assumption for $Y^{(n-\ell)}$ in the second sum of \eqn{step1}, which gives
\be
[XY]^{(n)}  \underset{p}{\rightarrow} \sum_{\ell = p+1}^{n-1}  \frac{1}{p}  \partial_i X^{(\ell - p)} \frac{\partial_i \theta^{(p)}}{\partial^2} Y^{(n - \ell)} + \sum_{\ell = 1}^{n-p-1} X^{(\ell)} \frac{1}{p} \partial_i Y^{(n-\ell - p)} \frac{\partial_i \theta^{(p)}}{\partial^2 } \ . 
\ee 
Then, relabeling the index in the first sum, this reduces to 
\be \label{xyprod}
[XY]^{(n)}  \underset{p}{\rightarrow} \frac{1}{p} \sum_{\ell = 1}^{n-p-1} \partial_i \left( X^{(\ell)}  Y^{(n-\ell - p)} \right) \frac{\partial_i \theta^{(p)}}{\partial^2 } = \frac{1}{p} \partial_i \left( [ X Y]^{(n-p)} \right) \frac{\partial_i \theta^{(p)}}{\partial^2 }  \ .
\ee
So, assuming the limits \eqn{xyirlimits} for all $p$ between $n-x$ and $n-1$, we have derived \eqn{xyprod} for $p = n - x -1$.  We will use this relationship shortly.

The other type of term we have is 
\be
\left[ \partial_i X \frac{\partial_i Y }{\partial^2} \right]^{(n)}  = \sum_{\ell = 1}^{n-1} \partial_i X^{(\ell)} \frac{\partial_i Y^{(n-\ell)} }{\partial^2}  \ . 
\ee
Similar to what we did above, we write out the sum to expose the terms that have $p$-particle soft limits
\be
\left[ \partial_i X \frac{\partial_i Y }{\partial^2} \right]^{(n)}  \underset{p}{\rightarrow} \sum_{\ell = p +1}^{n-1} \partial_i X^{(\ell)}\Big|_p  \frac{\partial_i Y^{(n-\ell)} }{\partial^2}  + \sum_{\ell = 1 }^{n-p} \partial_i X^{(\ell)} \frac{\partial_i Y^{(n-\ell)} }{\partial^2} \Big|_p  \ ,
\ee
where in the first sum, $ \partial_i X^{(\ell)}$ has the $p$-particle soft limits, and in the second sum $\partial_i Y^{(n - \ell )} / \partial^2$ has the $p$-particle soft limits (which now is non-zero even for $n - \ell = p$).  This can be rewritten as 
\be
\left[ \partial_i X \frac{\partial_i Y }{\partial^2} \right]^{(n)}  \underset{p}{\rightarrow} \partial_i X^{(n-p)} \frac{\partial_i Y^{(p)} }{\partial^2}  + \frac{1}{p} \sum_{\ell = 1}^{n - p - 1} \partial_i \left( \partial_j X^{(\ell)} \frac{\partial_j Y^{(n - p - \ell)}}{\partial^2} \right) \frac{\partial_i \theta^{(p)}}{\partial^2} \ , 
\ee
where we have relabeled one of the summation indices, and used the fact that the total momentum of $\partial_i \theta^{(p)} / \partial^2$ is going to zero to pull it out of derivatives.  This finally gives
\be \label{xyveloprod}
\left[ \partial_i X \frac{\partial_i Y }{\partial^2} \right]^{(n)}  \underset{p}{\rightarrow} \partial_i X^{(n-p)} \frac{\partial_i Y^{(p)} }{\partial^2}  + \frac{1}{p}  \partial_i \left(\left[ \partial_i X \frac{\partial_i Y }{\partial^2} \right]^{(n-p)}  \right) \frac{\partial_i \theta^{(p)}}{\partial^2} \ .
\ee

We are now ready to proceed with our induction.  We start with $p = n-1$.  This is the easiest case, since only the terms with $\partial_i \theta^{(n-1)} / \partial^2$ in \eqn{deltathetaab} contribute.  This gives
\begin{align}
\begin{split}
 & \delta^{(n)}  \underset{n-1}{\rightarrow} \frac{1}{(2 n +3)(n-1)}  \left[ ( 1 + 2n) \partial_i \delta^{(1)} \frac{\partial_i \theta^{(n-1)}}{\partial^2} + 2 \partial_i \theta^{(1)} \frac{\partial_i \theta^{(n-1)}}{\partial^2} \right] = \frac{1}{n-1} \partial_i \delta^{(1)} \frac{\partial_i \theta^{(n-1)}}{\partial^2} \ , \\
  & \theta^{(n)}  \underset{n-1}{\rightarrow} \frac{1}{(2 n +3)(n-1)}  \left[ 3 \partial_i \delta^{(1)} \frac{\partial_i \theta^{(n-1)}}{\partial^2}  + 2n  \partial_i \theta^{(1)} \frac{\partial_i \theta^{(n-1)}}{\partial^2} \right] = \frac{1}{n-1} \partial_i \theta^{(1)} \frac{\partial_i \theta^{(n-1)}}{\partial^2} \ , 
\end{split}
\end{align}
since $\delta^{(1)} = \theta^{(1)}$.  Now, we assume that \eqn{dmirlimits} is true for all $p$ from $n -1$ down to some $n - x$, for arbitrary $n$, and we show that this implies that \eqn{dmirlimits} is also satisfied for $p = n-x-1$.  So, let $p = n-x-1$ and take the $p$-particle soft limit of \eqn{deltathetaab}.  For that, we will need the $p$-particle soft limits of $A_n$ and $B_n$, which using \eqn{xyprod} and \eqn{xyveloprod}, we can immediately write as
\begin{align}
\begin{split}
& A_n \underset{p}{\rightarrow}  \partial_i \delta^{(n - p)} \frac{\partial_i \theta^{(p)} }{\partial^2} + \frac{1}{p} \partial_i A_{n-p} \frac{\partial_i \theta^{(p)} }{\partial^2}  \ ,  \\
& B_n \underset{p}{\rightarrow}  \partial_i \theta^{(n - p)} \frac{\partial_i \theta^{(p)} }{\partial^2} + \frac{1}{p} \partial_i B_{n-p} \frac{\partial_i \theta^{(p)} }{\partial^2}  \ .
\end{split}
\end{align}
Plugging this into \eqn{deltathetaab}, we find
\begin{align}
\begin{split}
\delta^{(n)} \underset{p}{\rightarrow} \frac{1}{(2 n +3)(n-1)} \frac{\partial_i \theta^{(p)}}{\partial^2} \partial_i \left[  (1 + 2 n ) \left(  \delta^{(n-p)} + \frac{A_{n-p}}{p} \right) + 2 \left(\theta^{(n-p)} + \frac{B_{n-p}}{p} \right) \right]  \ .
\end{split}
\end{align}
Solving for $A_{n-p}$ and $B_{n-p}$ in terms of $\delta^{(n-p)}$ and $\theta^{(n-p)}$ using \eqn{deltathetaab}, we find that the above expression simplifies to
\be \label{deltalim}
 \delta^{(n)} \underset{p}{\rightarrow} \frac{1}{p} \partial_i \delta^{(n-p)} \frac{\partial_i \theta^{(p)} }{\partial^2}  \ . 
\ee
Similarly, we have 
\begin{align}
\begin{split}
\theta^{(n)} \underset{p}{\rightarrow} \frac{1}{(2 n +3)(n-1)} \frac{\partial_i \theta^{(p)}}{\partial^2} \partial_i \left[  3  \left(  \delta^{(n-p)} + \frac{A_{n-p}}{p} \right) + 2 n  \left(\theta^{(n-p)} + \frac{B_{n-p}}{p} \right) \right]  \ ,
\end{split}
\end{align}
which simplifies to 
\be \label{thetalim}
 \theta ^{(n)} \underset{p}{\rightarrow} \frac{1}{p} \partial_i \theta^{(n-p)} \frac{\partial_i \theta^{(p)} }{\partial^2}  \ . 
\ee
As described above, we have only needed to use the induction assumption for $p$ from $n -1$ down to some $n - x$, and we have shown that this implies \eqn{dmirlimits} for $p = n - x - 1$, which completes the induction, and so we have proven \eqn{dmirlimits} for $1 \leq p \leq n-1$.

%
%
\subsection{Dark-matter $p$-particle soft limits in fluid basis} \label{dmplimfluidapp}

As shown in \cite{Donath:2023sav}, we can write the dark-matter overdensity as
\be \label{deltafl}
\delta^{(n)} = \sum_{\alpha = 1}^{n} \mathbb{C}_{\delta , \alpha}^{(n)}  \ , 
\ee
where for $\alpha < n$, we have the recursion relation
\be \label{singlefluiddeltarec}
\mathbb{C}_{\delta , \alpha}^{(n)} = \frac{1}{n-\alpha} \sum_{\ell = \alpha}^{n-1} \partial_i \mathbb{C}_{\delta , \alpha}^{(\ell)} \frac{\partial_i \theta^{(n-\ell)}}{\partial^2} \ .
\ee
In this appendix, we will show that $\mathbb{C}_{\delta , n}^{(n)}$ (i.e. with $\alpha = n$) has no $p$-particle soft limits, i.e.
\be
\mathbb{C}_{\delta , n}^{(n)} \underset{p}{\rightarrow}  0  \ ,
\ee
for all $n$ and $p$, which we will use later in \appref{multiplim}.  We will show this explicitly for $\delta$, from which the proof for $\theta$ can be trivially obtained by replacing $\delta \rightarrow \theta$.  Our strategy will be to show that the terms in the sum \eqn{deltafl} with $\alpha = 1 , \dots , n-1$ already satisfy the $p$-particle soft limits derived in the previous section, so that the limits of $\mathbb{C}_{\delta , n}^{(n)} $ must be trivial.

To do this, we will have to do an induction inside of an induction.  Ultimately, we will assume that $\mathbb{C}_{\delta , n'}^{(n')}$ has no $p$-particle soft limits for $1 \leq n' \leq n-1$, and show that this implies that $\mathbb{C}_{\delta , n}^{(n)}$ has no $p$-particle soft limits.  The base case $n'=1$ is $\mathbb{C}_{\delta , 1}^{(1)} = \delta^{(1)}$, which obviously has no $p$-particle soft limits.   Now, we start by assuming that $\mathbb{C}_{\delta , n'}^{(n')}$ has no $p$-particle soft limits for $1 \leq n' \leq n-1$.  Then we take the $p$-particle soft limit of \eqn{deltafl}, for $1 \leq p \leq n-1$.  First we write 
\be \label{deltafl2}
\delta^{(n)} = \mathbb{C}_{\delta , n}^{(n)} + \sum_{\alpha = 1}^{n-1} \mathbb{C}_{\delta , \alpha}^{(n)}  \ , 
\ee
and focus on the sum, which we will show already satisfies all of the limits of the previous section, so that $\mathbb{C}_{\delta , n}^{(n)}$ must have no $p$-particle soft limits.  So we now consider the kernels $\mathbb{C}_{\delta , \alpha}^{(n)}$ with $\alpha < n$.  

First, we notice that, for $n' < n$, 
\be \label{climexamp1}
\mathbb{C}_{\delta , \alpha}^{(n')}  \underset{p}{\rightarrow}  0 \ , \quad \text{for} \quad n' - \alpha < p \ . 
\ee
 To see this, consider \eqn{singlefluiddeltarec} for fixed $\alpha$ and $n'< n$, and find the highest $p$ for which $\mathbb{C}_{\delta , \alpha}^{(n')} $ has a $p$-particle soft limit.  Considering that $\alpha \leq n'$ in \eqn{climexamp1}, starting with $n' = \alpha + 1$, we have
\be
\mathbb{C}_{\delta , \alpha}^{(\alpha + 1)} = \partial_i \mathbb{C}_{\delta , \alpha}^{(\alpha)} \frac{\partial_i \theta^{(1)}}{\partial^2} \ , 
\ee
which, since $\mathbb{C}_{\delta, \alpha}^{(\alpha)}$ has no $p$-particle soft limit by the induction hypothesis, only has a non-zero limit for $p = 1$.  Thus the limit is zero for $ p>1$, satisfying \eqn{climexamp1}.  Next, consider $n' = \alpha + 2$, for which we have
\be
\mathbb{C}_{\delta , \alpha}^{(\alpha + 2)} = \frac{1}{2} \left( \partial_i \mathbb{C}_{\delta , \alpha}^{(\alpha)} \frac{\partial_i \theta^{(2)}}{\partial^2} + \partial_i \mathbb{C}_{\delta , \alpha}^{(\alpha+1)} \frac{\partial_i \theta^{(1)}}{\partial^2}   \right)  \ .
\ee
Since $\mathbb{C}_{\delta, \alpha}^{(\alpha)}$ has no $p$-particle soft limit, and $\mathbb{C}_{\delta, \alpha}^{(\alpha+1)}$ has no $p$-particle soft limit for $p >1$, $\mathbb{C}_{\delta , \alpha}^{(\alpha + 2)}$ only has a non-zero $p$-particle soft limit for $p = 1$ and $p =2$ (coming from $\partial_i \theta^{(2)} / \partial^2$), and so has a zero limit for $p > 2$, also satisfying \eqn{climexamp1}.  This trend continues, since $\mathbb{C}_{\delta, \alpha}^{(\alpha+s)}$ only involves $\mathbb{C}_{\delta, \alpha}^{(\ell)}$ for $\alpha \leq \ell \leq \alpha+ s -1$ and  $\partial \theta^{(r)} / \partial^2$ for $ 1 \leq r \leq s$, and for these ranges of $\ell$ and $r$, $\mathbb{C}_{\delta, \alpha}^{(\ell)}$ has zero limits for $p > s-1$ and $\partial \theta^{(r)} / \partial^2$ has zero limits for $p >s$.  Thus,  $\mathbb{C}_{\delta, \alpha}^{(\alpha+s)}$ has zero limits for $p >s$, which is equivalent to $p > n' - \alpha$ with $n' = \alpha + s$, thus proving \eqn{climexamp1}.

Now we want to show that 
\be \label{ckerplim}
\mathbb{C}_{\delta , \alpha}^{(n)} \underset{p}{\rightarrow} \frac{1}{p} \partial_i \mathbb{C}_{\delta , \alpha}^{(n-p)}  \frac{\partial_i \theta^{(p)}}{\partial^2} \ , 
\ee
for $1 \leq \alpha \leq n-1$ and $1 \leq p \leq n- \alpha$ (as for larger $p$, the limits are zero by \eqn{climexamp1}).  To prove \eqn{ckerplim}, we have to do another induction on $p$, similar to the one done in the previous section, starting from $p = n - \alpha$ and moving down in $p$.  Using \eqn{singlefluiddeltarec}, we see that the base case of $p = n-\alpha$ is easy, since only the $\ell = \alpha$ term contributes, as because of \eqn{climexamp1} all of the $\mathbb{C}_{\delta, \alpha}^{(\ell)}$ in the sum have zero limit, and we obtain
\be
\mathbb{C}_{\delta , \alpha}^{(n)} \underset{n-\alpha }{\rightarrow} \frac{1}{p} \partial_i \mathbb{C}_{\delta , \alpha}^{(n-p)} \frac{\partial_i \theta^{(p)}}{\partial^2}  \ , \quad \text{for} \quad p = n - \alpha \ . 
\ee
Now we assume that \eqn{ckerplim} holds for all $p$ from $n - \alpha$ down to $n - \alpha - x$ for arbitrary $n$, and we show that this implies that \eqn{ckerplim} holds for $p = n - \alpha - x - 1$.  We take the $p$-particle soft limit of \eqn{singlefluiddeltarec}, and use similar manipulations to those that lead to \eqn{xyveloprod}.  Specifically, we have for $p = n - \alpha - x - 1$, 
\begin{align}
\begin{split} \label{cdeltaalphaplim}
\mathbb{C}_{\delta , \alpha}^{(n)}&  \underset{p }{\rightarrow} \frac{1}{n-\alpha} \sum_{\ell = \alpha}^{n-1} \left(  \partial_i \mathbb{C}_{\delta , \alpha}^{(\ell)} \Big|_p \frac{\partial_i \theta^{(n-\ell)}}{\partial^2}  +  \partial_i \mathbb{C}_{\delta , \alpha}^{(\ell)} \frac{\partial_i \theta^{(n-\ell)}}{\partial^2}\Big|_p  \right) \\
& =\frac{1}{n-\alpha}  \left( \sum_{\ell = \alpha+p}^{n-1} \partial_i \mathbb{C}_{\delta , \alpha}^{(\ell)} \Big|_p \frac{\partial_i \theta^{(n-\ell)}}{\partial^2}  + \sum_{\ell = \alpha}^{n-p}  \partial_i \mathbb{C}_{\delta , \alpha}^{(\ell)} \frac{\partial_i \theta^{(n-\ell)}}{\partial^2}\Big|_p  \right) \\
& =\frac{1}{n-\alpha}  \left( \sum_{\ell = \alpha+p}^{n-1} \partial_i \mathbb{C}_{\delta , \alpha}^{(\ell)} \Big|_p \frac{\partial_i \theta^{(n-\ell)}}{\partial^2}  +  \partial_i \mathbb{C}_{\delta , \alpha}^{(n-p)} \frac{\partial_i \theta^{(p)}}{\partial^2} + \sum_{\ell = \alpha}^{n-p-1}  \partial_i \mathbb{C}_{\delta , \alpha}^{(\ell)} \frac{\partial_i \theta^{(n-\ell)}}{\partial^2}\Big|_p  \right) \\
& =\frac{1}{n-\alpha}  \Bigg( \partial_i \mathbb{C}_{\delta , \alpha}^{(n-p)} \frac{\partial_i \theta^{(p)}}{\partial^2} + \sum_{\ell = \alpha+p}^{n-1} \partial_i \left(\frac{1}{p} \partial_j \mathbb{C}_{\delta, \alpha}^{(\ell - p)} \frac{\partial_j \theta^{(p)}}{\partial^2}  \right) \frac{\partial_i \theta^{(n-\ell)}}{\partial^2}   \\
& \hspace{1in} + \sum_{\ell = \alpha}^{n-p-1}  \partial_i \mathbb{C}_{\delta , \alpha}^{(\ell)} \frac{\partial_i}{\partial^2}\left( \frac{1}{p} \partial_j \theta^{(n - \ell - p)} \frac{\partial_j \theta^{(p)}}{\partial^2} \right)  \Bigg) \\
& = \frac{1}{n - \alpha}  \left(  \partial_i \mathbb{C}_{\delta , \alpha}^{(n - p)} \frac{\partial_i \theta^{(p)}}{\partial^2} + \frac{1}{p} \sum_{\ell = \alpha}^{n - p -1} \partial_i \left( \partial_j \mathbb{C}_{\delta , \alpha}^{(\ell)} \frac{\partial_j \theta^{(n - p - \ell)}}{\partial^2}   \right) \frac{\partial_i \theta^{(p)}}{\partial^2} \right)  \\
& = \frac{1}{n - \alpha}  \left(  \partial_i \mathbb{C}_{\delta , \alpha}^{(n - p)} \frac{\partial_i \theta^{(p)}}{\partial^2} + \frac{n-p-\alpha}{p}  \partial_i \mathbb{C}_{\delta , \alpha}^{(n-p)} \frac{\partial_i \theta^{(p)}}{\partial^2} \right) \\
& = \frac{1}{p} \partial_i \mathbb{C}_{\delta , \alpha}^{(n-p)} \frac{\partial_i \theta^{(p)}}{\partial^2}  \ ,
\end{split}
\end{align}
where from the first to second line we limited the sums to the terms that have non-zero limits (using \eqn{climexamp1} and the results for $\theta$ in \secref{dmplimapp} that say that $\partial_i \theta^{(n-\ell)} / \partial^2$ has a non-zero $p$-particle limit for $n-\ell \geq p$), from the second to third line we separated the term with $\ell = n-p$ from the second sum, from the third to fourth line we plugged in the $p$-particle limit of $\theta$ \eqn{dmirlimits} and the induction assumption for the $p$-particle limit of $\mathbb{C}_{\delta , \alpha}^{(\ell)}$ \eqn{ckerplim},\footnote{ The induction assumption for $\mathbb{C}_{\delta , \alpha}^{(\ell)}$ applies for $  \ell - \alpha - x \leq p \leq n - \alpha$, which for the top limit of the sum $\ell = n -1$, means it applies to that term for $  n-1-\alpha -x \leq p \leq n - \alpha$, and $p = n - 1 - \alpha - x$ is the limit we are taking, so we can use the induction assumption. For the lower $\ell$ in the sum, it clearly applies as well.} from the fourth to the fifth line we shifted summation variables in the first sum and pulled $\partial_i \theta^{(p)} / \partial^2$ out of the derivatives, from the fifth to the sixth line we used the recursion relation \eqn{singlefluiddeltarec}, and from the sixth to the seventh line we simply added the terms.   Thus, we successfully complete the induction and prove \eqn{ckerplim}.  

Going back to \eqn{deltafl2}, this now means that
\be
\delta^{(n)} - \mathbb{C}_{\delta , n}^{(n)}  \underset{p }{\rightarrow} \sum_{\alpha = 1}^{n-p} \frac{1}{p}   \partial_i \mathbb{C}_{\delta , \alpha}^{(n-p)}  \frac{\partial_i \theta^{(p)}}{\partial^2} = \frac{1}{p} \partial_i \delta^{(n-p)} \frac{\partial_i \theta^{(p)}}{\partial^2}  \ , 
\ee
and since $\delta^{(n)}$ on the left-hand side already satisfies this \eqn{dmirlimits}, we conclude that $\mathbb{C}_{\delta , n}^{(n)}$ must have no $p$-particle soft limits.  This completes the final induction.

%
%
\subsection{General non-local-in-time $p$-particle soft limits} \label{multiplim}

In this section, we ultimately want to show that 
\be \label{galppartapp}
\delta^{(n)}_g \underset{p }{\rightarrow}  \frac{1}{p} \partial_i \delta^{(n-p)}_g \frac{\partial_i \theta^{(p)}}{\partial^2} \ ,
\ee
for $1 \leq p \leq n-1$.  We will do this by starting with \eqn{finalfulltauij} and writing it as
\be \label{deltagsumexp}
\delta^{(n)}_g = \sum_{m =1}^{n} \sum_{\{ \mathcal{O}_m\}} \sum_{\alpha = m}^{n} \sum_{\{\alpha_1 , \dots , \alpha_m\}_{\alpha}} c_{\mathcal{O}_m; \alpha_1 , \dots , \alpha_m }   \bar{ \mathbb{C}}^{(n)}_{\mathcal{O}_m; \alpha_1 , \dots , \alpha_m}  \ , 
\ee
where $\{\alpha_1 , \dots , \alpha_m\}_{\alpha}$ is the set of all tuples $(\alpha_1 , \dots , \alpha_m)$ with $\alpha_1 + \cdots + \alpha_m = \alpha$.  Using \eqn{finrec}, we can see that 
\be
\bar{\mathbb{C}}^{(n)}_{\mathcal{O}_m; \alpha_1 , \dots , \alpha_m} \Big|_{\sum_{i =1 }^m \alpha_i  =  n} \underset{p }{\rightarrow}  0 \ ,
\ee
because it is a product of factors from the $\delta$ or $\theta$ fluid expansion with $n = \alpha$, and in the previous section we showed that they all have no $p$-particle soft limits.  Thus, we can focus on the terms of \eqn{deltagsumexp} with $m \leq \sum_{i =1 }^m \alpha_i  \leq n-1$, which we will now show satisfy
\be \label{cbarlimit}
\bar{ \mathbb{C}}^{(n)}_{\mathcal{O}_m; \alpha_1 , \dots , \alpha_m}  \underset{p }{\rightarrow}  \frac{1}{p}  \partial_i  \bar{ \mathbb{C}}^{(n-p)}_{\mathcal{O}_m; \alpha_1 , \dots , \alpha_m} \frac{\partial_i \theta^{(p)}}{\partial^2} \ ,
\ee 
for $1 \leq p \leq n - \sum_{i=1}^m \alpha_i $.  

First, we notice that the recursion relation for $\bar{ \mathbb{C}}^{(n)}_{\mathcal{O}_m; \alpha_1 , \dots , \alpha_m} $ in \eqn{newxflrec} is exactly the same form as the recursion relation for $\mathbb{C}_{\delta , \alpha}^{(n)} $ in  \eqn{singlefluiddeltarec} identifying $\alpha \rightarrow \sum_{i=1}^m \alpha_i $.  The only difference is that $\sum_{i=1}^m \alpha_i \geq m$, while in \eqn{singlefluiddeltarec}, $\alpha \geq 1$.   Thus, anything we derived from the recursion relation \eqn{singlefluiddeltarec} for $\mathbb{C}_{\delta , \alpha}^{(n)} $, we can analogously derive from the recursion relation \eqn{newxflrec} for $\bar{ \mathbb{C}}^{(n)}_{\mathcal{O}_m; \alpha_1 , \dots , \alpha_m} $, as long as the proof did not depend on the value of $\alpha$.  Thus, we immediately have 
\be \label{cnalphapzero}
\bar{ \mathbb{C}}^{(n)}_{\mathcal{O}_m; \alpha_1 , \dots , \alpha_m}   \underset{p}{\rightarrow}  0 \ , \quad \text{for} \quad n - \sum_{i = 1}^{m} \alpha_i  < p \ ,
\ee
since the proof of \eqn{climexamp1} did not depend on the value of $\alpha$.  Next, using this, we can run through the exact manipulations as in \eqn{cdeltaalphaplim} (but identifying $\alpha \rightarrow \sum_{i=1}^m \alpha_i $), which also did not depend on the value of $\alpha$. Thus, we have proven \eqn{cbarlimit}.

Given the results above, we can now take the $p$-particle limit of $\delta_g^{(n)}$ to get 
\begin{align}
\begin{split}
\delta^{(n)}_g & \underset{p }{\rightarrow} \sum_{m =1}^{n-p} \sum_{\{ \mathcal{O}_m\}} \sum_{\alpha = m}^{n-p} \sum_{\{\alpha_1 , \dots , \alpha_m\}_{\alpha}} c_{\mathcal{O}_m; \alpha_1 , \dots , \alpha_m }   \frac{1}{p}  \partial_i  \bar{ \mathbb{C}}^{(n-p)}_{\mathcal{O}_m; \alpha_1 , \dots , \alpha_m} \frac{\partial_i \theta^{(p)}}{\partial^2} \\
& =  \frac{1}{p}   \partial_i  \left(   \sum_{m =1}^{n-p} \sum_{\{ \mathcal{O}_m\}} \sum_{\alpha = m}^{n-p} \sum_{\{\alpha_1 , \dots , \alpha_m\}_{\alpha}} c_{\mathcal{O}_m; \alpha_1 , \dots , \alpha_m }  \bar{ \mathbb{C}}^{(n-p)}_{\mathcal{O}_m; \alpha_1 , \dots , \alpha_m} \right) \frac{\partial_i \theta^{(p)}}{\partial^2} \\
& = \frac{1}{p} \partial_i \delta_g^{(n-p)} \frac{\partial_i \theta^{(p)}}{\partial^2} \ .
\end{split}
\end{align}
  In the first line, the sum over $\alpha$ goes from $m$ to $n -p$ because of \eqn{cnalphapzero}, and to go to the third line, we used \eqn{deltagsumexp} with $n \rightarrow n - p$.  Thus, we have proven \eqn{galppartapp} for $1 \leq p \leq n-1$.

\end{appendix}

%
%
%
%

 \bibliographystyle{JHEP}
 \small
\bibliography{references_3}

 \end{document}